\documentclass[sn-mathphys]{sn-jnl_edited}% Math and Physical Sciences Reference Style
\jyear{2023}%

\theoremstyle{thmstyleone}%
\theoremstyle{thmstyletwo}%

\theoremstyle{thmstylethree}%

\usepackage{bm}

\usepackage{siunitx} % voor het gebruik van eenheden

\usepackage{pdflscape}
\usepackage{gensymb}

\usepackage{listings}

\definecolor{codegreen}{rgb}{0,0.6,0}
\definecolor{codegray}{rgb}{0.5,0.5,0.5}
\definecolor{codepurple}{rgb}{0.58,0,0.82}
\definecolor{backcolour}{rgb}{0.95,0.95,0.92}

\lstdefinestyle{mystyle}{
    backgroundcolor=\color{backcolour},   
    commentstyle=\color{codegreen},
    keywordstyle=\color{magenta},
    numberstyle=\tiny\color{codegray},
    stringstyle=\color{codepurple},
    basicstyle=\ttfamily\footnotesize,
    breakatwhitespace=false,         
    breaklines=true,                 
    captionpos=b,                    
    keepspaces=true,                 
    numbers=left,                    
    numbersep=5pt,                  
    showspaces=false,                
    showstringspaces=false,
    showtabs=false,                  
    tabsize=2
}

\usepackage{mhchem}

\graphicspath{{fig/}}

\usepackage{titlesec}
\titleformat{\paragraph}[runin]{\normalfont\bfseries}{\theparagraph}{}{}[ ~ ]

\begin{document}

\title{Validating a dynamic input-output model for the propagation of supply and demand shocks during the COVID-19 pandemic in Belgium}


\author*[1]{\fnm{Tijs W.}
\sur{Alleman}}\email{tijs.alleman@ugent.be}

\author[2]{\fnm{Koen}
\sur{Schoors}}

\author[1]{\fnm{Jan M.}
\sur{Baetens}}

\affil[1]{\orgdiv{BionamiX}, \orgname{Department of Data Analysis and Mathematical Modelling, Ghent University}, \orgaddress{\street{Coupure Links 653}, \city{Ghent}, \postcode{9000}, \country{Belgium}}}

\affil[2]{\orgdiv{Complex Systems Institute (CSI)}, \orgname{Department of Economics, Ghent University}, \orgaddress{\street{Tweekerkenstraat 2}, \city{Ghent}, \postcode{9000}, \country{Belgium}}}


\maketitle

{\small
\noindent\textbf{Abstract} This work validates a dynamic production network model, used to quantify the impact of economic shocks caused by COVID-19 in the UK, using data for Belgium. Because the model was published early during the 2020 COVID-19 pandemic, it relied on several assumptions regarding the magnitude of the observed economic shocks, for which more accurate data have become available in the meantime. We refined the propagated shocks to align with observed data collected during the pandemic and calibrated some less well-informed parameters using 115 economic time series. The refined model effectively captures the evolution of GDP, revenue, and employment during the COVID-19 pandemic in Belgium at both individual economic activity and aggregate levels. However, the reduction in business-to-business demand is overestimated, revealing structural shortcomings in accounting for businesses' motivations to sustain trade despite the pandemic's induced shocks. We confirm that the relaxation of the stringent Leontief production function by a survey on the criticality of inputs significantly improved the model’s accuracy. However, despite a large dataset, distinguishing between varying degrees of relaxation proved challenging. Overall, this work demonstrates the model’s validity in assessing the impact of economic shocks caused by an epidemic in Belgium.\\


%\noindent\textbf{Word count} xxxx words (main text) excluding captions of figures and tables.\\
}

%%%%%%%%%%%%%%%%%%%%%%%%%%%%%%%%%%%%%%%%%%%%%%%%%%%%%%%%%%%%%%%%%%%%%%%%%%%%%%
%%%%%%%%%%%%%%%%%%%%%%%%%%%%%%%%%%%%%%%%%%%%%%%%%%%%%%%%%%%%%%%%%%%%%%%%%%%%%%

\section{Introduction}

The COVID-19 pandemic and the measures imposed to counter the spread of SARS-CoV-2 created severe disruptions to economic output globally, as evidenced by the profound impact on the Belgian economy. In 2020, 11.3~\% of the employees were temporarily unemployed and were furloughed 70~\% of their salary, amounting to an estimated 16.3 billion EUR of government spending \citep{ermg2021,struyven2021}. Further, the support offered to Belgian firms in the form of allowances and tax exemptions has been exceptional at a total expenditure in 2020 and 2021 of 11.6 billion EUR \citep{piette2022}. In addition to economic disruption, the COVID-19 pandemic also inflicted numerous consequences on public health \citep{mehta2022,davis2023} and well-being \citep{reiriz2023,tiete2021,sun2023,butterworth2022}, and education \citep{lee2020,maldonado2020,cohen2022,maisonneuve2023}. During the COVID-19 pandemic, policymakers worldwide thus faced a considerable challenge in effectively managing the virus' spread while minimising the devastating impacts on public health, education and economic activity.\\

Within this context, mathematical models emerged as promising tools to assess the impact of government policies on pandemic progression and the impacts on public health, education and economic activity. Pichler et al. \cite{pichler2020} proposed a dynamic production network model (PNM) coupled to a disease transmission model (DTM) to investigate the epidemiological and economic impact of several lockdown release scenarios for the UK in May 2020. The model differs from traditional out-of-equilibrium PNMs in the way input bottlenecks are treated in the production function. Under the Leontief production function, every input is considered essential for production, and thus if a single input is unavailable, overall production will be reduced proportionally. As an example, the construction sector has restaurants as an input, most likely to entertain clients. Closing restaurants during the COVID-19 pandemic results in a sharp drop in the output of the construction sector under a Leontief production function, which is not realistic, especially in the short term. Using a survey on the criticality of inputs to production, Pichler et al. \cite{pichler2022} demonstrated how the stringent Leontief production function can be relaxed to a partially-binding Leontief (PBL) production function, resulting in a higher descriptive accuracy of empirical data.\\

Before building a coupled epidemiological-economic co-simulation model, we wish to demonstrate the validity of the dynamic PNM by Pichler et al. \citep{pichler2020,pichler2022} in a Belgian context. Because the model was published early during the 2020 COVID-19 pandemic, it relied on several assumptions regarding the magnitude of the observed economic shocks, for which more accurate data have become available in the mean time. We implement three important improvements. First, we use survey data gathered by the Economic Risk Management Group of the Belgian National Bank (ERMG) \cite{ermg2021} on the number of temporarily unemployed workers during the COVID-19 pandemic to infer the magnitude of the labor supply shock, as opposed to using the index proposed by del Rio-Chanona et al. \cite{delriochanona2020} based on the estimated ability to work from home and importance of an economic activity. Second, we assume the shocks to demand from government, non-profits and exports of services have the same magnitude and time course as the shocks to household demand, and the shock to investments and the exports of goods are brought in line with trade data observe during the 2020-2021 COVID-19 pandemic \citep{oecd2023,nbb2023b}, as opposed to having no shocks on government and non-profit demand and imposing a 15~\% shock to all investments and exports uniformly \citep{pichler2022}. Third, we calibrate the household demand shock under lockdown using 115 time series of aggregated and sectoral data, as opposed to using a ballpark figure proposed by the US Congressional Budget Office \cite{congressional_budget_office2006}. To this end, we use a total of 115 aggregated and sectoral time series of business-to-business (B2B) transactions, synthetic gross domestic product (GDP), revenue, and employment. After determining the optimal magnitude of the shocks, along with the optimal values of some other less well-informed parameters, the accuracy of the model in describing the empirical time series is assessed both at the aggregated and sectoral levels. Finally, we assess to what degree each of the aforementioned modifications contributes to greater descriptive accuracy.

\section{Methods}

\subsection{Dynamic production network model}
\subsubsection{Overview}

We adopt the dynamic PNM by Pichler et al. \cite{pichler2022}. Economic activity is classified in $N=63$ sectors corresponding to the \textit{Nomenclature des Activit\'es \'Economiques dans la Communaut\'e Europ\'eenne} (NACE) \citep{NACE}. An index of the aggregation of NACE in 21 economic activities (sectors) is given in Table \ref{tab:NACE21} and an aggregation of 63 economic activities is given in Table \ref{tab:NACE64}. The model uses the $63 \times 63$ input-output matrix of Belgium \citep{FPB2018} to inform the intermediate flows of services and products in the Belgian production network. The economy uses intermediates and labor to satisfy the demand of two end users: households and \textit{other sources} (government and non-profit consumption, investments, and exports). We model the changes in other demand during the COVID-19 pandemic exogenously using trade data for Belgium (Section \ref{section:EPNM_shocks}). The gross output of sector $i$ is the sum of the intermediate consumption of its goods by other companies, households, and others. Mathematically, its basic accounting structure is as follows,
\begin{equation}
    x_i(t) = \sum_{j=1}^N Z_{ij}(t) + c_{i}(t) + f_{i}(t)
\end{equation}
where $x_i(t)$ is the gross output of sector $i$. $Z_{ij}(t)$ is the input-output matrix containing the intermediate consumption of good $i$ by industry $j$. $c_i(t)$ is the household consumption of good $i$ and $f_i(t)$ is the other consumption of good $i$. We adopt the standard convention that in the input-output matrix columns represent demand while rows represent supply. Prices are assumed time-invariant and capital is not explicitly modeled. One representative firm is modeled for each sector and there is one representative household. Every firm keeps an inventory of inputs from all other firms and draws from these inventories to produce outputs. Intermediates in production are modeled as deliveries replenishing the firm's inventory. The model tracks the dynamics of seven relevant variables such as gross output and labor compensation (Table \ref{tab:EPNM_overview_states}). Prior to the COVID-19 pandemic, the economy is in equilibrium and supply equals demand. The pandemic imbalances the economy through a combination of shocks in household demand, other demand, and labor supply (Section \ref{section:EPNM_shocks}). Further, firms may run out of intermediate inputs and may need to stop production. However, as opposed to a traditional Leontief production function, not every intermediate input may be critical to production \citep{pichler2022}. A schematic overview of the model is shown in Figure \ref{fig:economic_model}, while  its parameters and their values are listed in Table \ref{tab:overview_parameters}. At each timestep $t$ the model loops through the following steps (Fig. \ref{fig:economic_model}).
\begin{enumerate}\itemsep0.5em 
    \item The values of the household demand shock $\kappa_i^D(t)$, other demand shock $\kappa_i^F(t)$, labor supply shock $\kappa_i^S(t)$, and, the reimbursed fraction of workers' lost salary $b(t)$, which are the model's inputs, are retrieved (Fig.~\ref{fig:economic_model}).
    \item Total desired demand, which is the sum of desired household demand, other demand and B2B demand are computed subject to aforementioned shocks.
    \item Firms will produce as much as they can to satisfy demand, thus the maximum productive capacity under constrained labor availability and under available inputs is computed. In line with Pichler et al. \cite{pichler2022}, input bottlenecks are treated in five different ways depending on the criticality of the inputs.
    \item The realized output is computed. If the realized output does not meet demand, then industries ration their output proportionally across households, exogenous agents, and businesses.
    \item The inventories of each firm are updated using the realized B2B demand.
    \item Firms hire or fire workers depending on their ability to meet demand.
    \item The system is integrated continuously using the \textit{Runge-Kutta 45} algorithm or discretely with a step size of up to one day (Figure \ref{fig:impact_discrete_continuous}). Both methods are available through pySODM \citep{alleman2023b}.
\end{enumerate}

\begin{table}[!h]
    \centering
    \caption[Overview of the production network model's states.]{Overview of model states. The initial values of the states are listed in Table \ref{tab:overview_initial_states}.}
    {\footnotesize\renewcommand{\arraystretch}{1.10}
    \begin{tabular}{p{1.2cm}p{8.0cm}}
        \toprule
        \textbf{Symbol} & \textbf{Name} \\ \midrule
        $x_{i}(t)$ & Gross output of sector $i$ at time $t$\\
        $d_{i}(t)$ & Total demand of sector $i$ at time $t$\\
        $l_{i}(t)$ & Labor compensation to workers in sector $i$ at time $t$\\
        $c_{i}(t)$ & Realised household demand of good $i$ at time $t$\\
        $f_{i}(t)$ & Realised other demand of good $i$ at time $t$\\
        $O_{ij}(t)$ & Realised B2B demand by sector $i$ of good $j$ at time $t$\\
        $S_{ij}(t)$ & Stock of material $i$ held in the inventory of sector $j$ at time $t$\\
      \bottomrule
    \end{tabular}
    }
    \label{tab:EPNM_overview_states}
\end{table}

\subsubsection{Government measures and economic shocks}\label{section:EPNM_shocks}

\begin{figure}[bh!]
    \centering
    \includegraphics[width=1\linewidth]{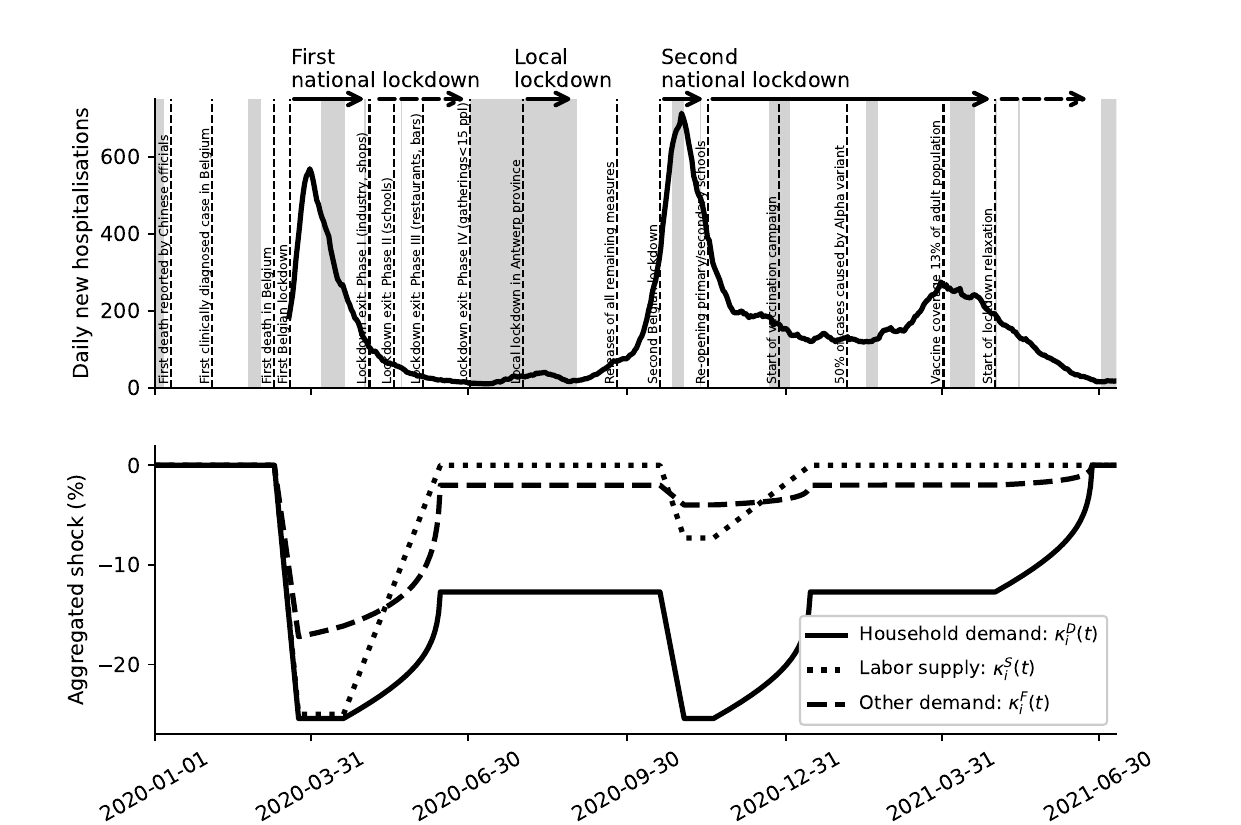}
    \caption[Timeline of the 2020-2021 pandemic in Belgium.]{2020-2021 pandemic timeline: (top) Major COVID-19-related events (dashed vertical lines). 7-day moving average of daily new COVID-19 hospitalisations in Belgium \citep{sciensano2023} (solid line). The solid horizontal arrows denote periods with severe restrictions, the dashed horizontal arrows denote the release of these restrictions. A grey background represents the school holidays. Adapted and extended from \cite{rollier2023}. (bottom) Aggregated consumer demand shock (solid line), aggregated other demand shock (dotted line), and labor supply shock (dashed line).} 
    \label{fig:timeline}
\end{figure}

\paragraph{Pandemic timeline} During the 2020-2021 pandemic, lockdown measures were taken twice in Belgium, first on March 15, 2020, and then again on October 19, 2020 (Fig.~\ref{fig:timeline}). The first lockdown lasted eight weeks, from March 15, 2020 until May 4, 2020, and involved the mandatory closure of schools, a ban on all forms of leisure activities (accommodation, rental, travel, recreation), the closure of all non-essential retail, a travel ban, and mandated work-at-home wherever distancing could not be guaranteed \citep{luyten2021}. Given the complete absence of personal protective equipment, this amounted to a \textit{de facto} closure of most economic activities. Starting May 4, 2020, lockdown measures were eased in a step-wise fashion, and by July 1, 2020, most of the restrictions were lifted. During July 2020, COVID-19 incidence started rising again prompting a local lockdown in Antwerp province during August 2020. After a summer marked by low COVID-19 incidence, the situation deteriorated quickly during September 2020 and on October 19, 2020, the Belgian government was forced to impose lockdown measures for a second time. The measures taken were very similar to the first lockdown, however, a key difference is that work-at-home was only mandated for those who could work from home, resulting in no forced economic shutdowns (besides those in accommodation, rental, travel, recreation). On November 16, 2020, elementary and high schools reopened, but all other restrictions stayed in place until mid-May 2021. This lockdown, where economic activity was permitted but leisurely activities were not, was colloquially referred to as \textit{lockdown light}. By July 2021, practically all measures were lifted.\\

\begin{table}[!th]
    \centering
    \caption[Overview of the production network model's parameters.]{Overview of model parameters. $\mathcal{U}(\text{lb},\ \text{ub})$ is used to denote a uniform distribution while $\mathcal{N}(\mu, \sigma^2)$ is used to denote a normal distribution.}
    {\footnotesize\renewcommand{\arraystretch}{1.10}
    \begin{tabular}{p{0.6cm}p{5.2cm}p{2.2cm}p{2.0cm}}
        \toprule
        \textbf{Symb.} & \textbf{Name} & \textbf{Value} & \textbf{Source} \\ \midrule
        $\bm{\kappa}^D$ & Elements $\kappa_i^D$. Household demand shock to sector $i$ under lockdown. & Table \ref{tab:overview_shocks} & Section \ref{section:EPNM_results} \\
        $\bm{\kappa}^F$ & Elements $\kappa_i^F$. Other demand shock to sector $i$ under lockdown. & Table \ref{tab:overview_shocks} & Section \ref{section:EPNM_results} \\
        $\bm{\kappa}^S$ & Elements $\kappa_i^S$. Labor supply shock to sector $i$ under lockdown. & Table \ref{tab:overview_shocks} & Section \ref{section:EPNM_results} \\
        $b$ & Reimbursed fraction of lost labor income & 0.7 & \cite{thuy2020}\\
        \midrule
        $\bm{Z}$ & Elements $Z_{ij}$. Intermediate consumption by sector $i$ of good $j$. Input-Output matrix. & -- & \cite{FPB2018}\\
        $\bm{\mathcal{A}}$ & Elements $\mathcal{A}_{ij}$. Technical coefficients. Payment to sector $i$ per unit produced of $j$. & $\mathcal{A}_{ij} = Z_{ij}/x_{j}(0)$ & Computed\\
        $\bm{\mathcal{C}}$ & Elements $\mathcal{C}_{ij}$. Critical inputs $j$ of sector $i$ & Survey & \cite{pichler2022} \\
        $\bm{\mathcal{I}}$ & Elements $\mathcal{I}_{ij}$. Important inputs $j$ of sector $i$ & Survey & \cite{pichler2022} \\
        $\bm{n}$ & Elements $n_j$. Targeted number of days inventory of sector $j$ by sector $i$ & Table \ref{tab:overview_initial_states} & \cite{pichler2022} \\  
        $\mathcal{F}$ & Production function & Half-critical & Section \ref{section:EPNM_results}\\
        $\tau$ & Speed of inventory restocking & $\mathcal{N}(1, 0.2)~\text{days}$ & Section \ref{section:EPNM_results} \\     
        $\iota_F$ & Speed of firing & $\mathcal{N}(28, 7)~\text{days}$ & Section \ref{section:EPNM_results} \\
        $\iota_H$ & Speed of hiring & $\iota_H=2*\iota_F$  & \cite{pichler2022} \\
        $\rho$ & Aggregate household consumption adjustment speed & $\mathcal{U}(0.5,1)~\text{quarters}$ & Assumption, \cite{muellbauer2020} \\
        $\Delta s $ & Changes in the savings rate & $\mathcal{U}(0.5,1)$ & \cite{basselier2021}\\
        $m$ & Share of labor income used to consume final domestic goods & 0.86 & Computed\\
        $L$ & Fraction of households believing in an L-shaped economic recovery & 1 & \cite{nbb2021}\\
        $l_1$ & Number of days needed to ease in shocks & $\mathcal{N}(14,2)~\text{days}$ & Assumption  \\
        $l_2$ & Number of days needed to ease out shocks after lockdown & $\mathcal{N}(56,7)~\text{days}$ & Assumption \\
        $r$ & Fraction of household and other demand shock under lockdown, remaining between both lockdowns and during the \textit{lockdown light} & $\mathcal{U}(0.25,0.75)$ & Section \ref{section:EPNM_results}  \\
      \bottomrule
    \end{tabular}
    }
    \label{tab:overview_parameters}
\end{table}

\begin{landscape}
\thispagestyle{empty}
\begin{figure}[h!]
    \centering
    \includegraphics[width=0.99\linewidth]{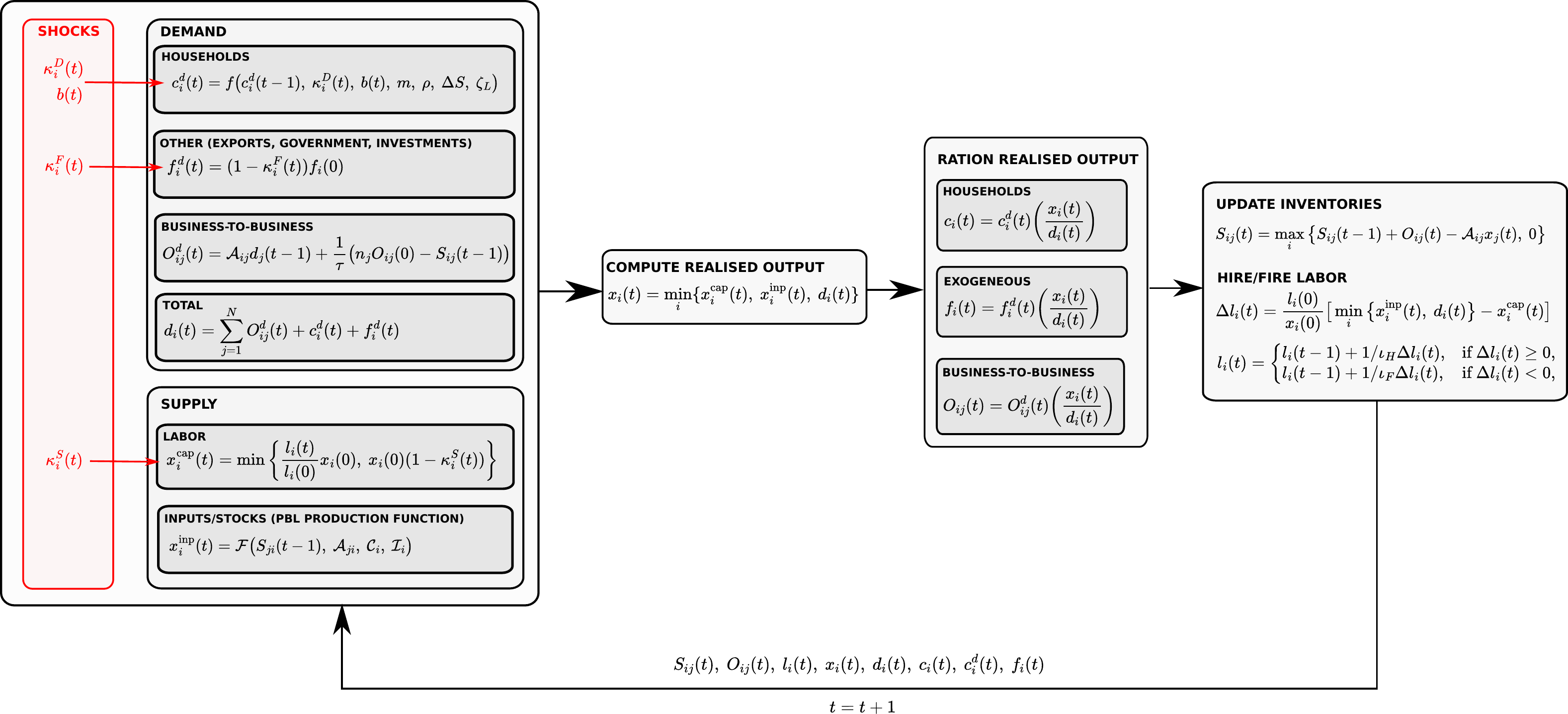}
    \caption[Schematic representation of the dynamic production network model.]{Schematic representation of the dynamic PNM by \cite{pichler2022}.} 
    \label{fig:economic_model}
\end{figure}
\end{landscape}

\paragraph{Household demand shocks} During lockdown, households reduce their demand for customer-contact services, such as restaurants, either due to fear of infection or prohibition of consumption, mathematically denoted as $\kappa_{i}^D$. Pichler et al. \cite{pichler2022} based their household demand shock on a study on the economic impact of pandemic influenza by the US Congressional Budget Office \cite{congressional_budget_office2006} (Table \ref{tab:overview_shocks_pichler}). In this study, the largest consumer demand shock of $-80~\%$ is experienced by the Accommodation \& Food service (I55-56), Rental and Leasing (N77), Travel agencies (N79), Recreation (R90-91-92-93) and the Activities of Membership Organisations (S94) and Other Personal Service Activities (S96), while Land, water, and air transport (H49-50-51) undergo a moderate demand shock of $-67~\%$. The Agricultural (A), Mining (B), Manufacturing (C), and Wholesale and Retail (G) undergo minor consumer demand shocks of -10~\%. The remaining sectors face no consumer demand shocks (Table \ref{tab:overview_shocks_pichler}). In this work, we will validate the magnitude of these shocks, which were ballpark figures, by calibrating them to empirical data (Section \ref{section:sensitivity_analysis}). After lockdown, we assume the shocks to household demand for sectors involving on-site consumption (Table \ref{tab:overview_shocks_pichler}) are eased using the following non-linear function,
\begin{equation}\label{eq:EPNM_timecourse_shocks}
    \kappa_{i}^D(t) = 
    \begin{cases}
        \kappa_{i}^D, & \text{if } t_{\text{start}} \leq t < t_{\text{end}} \\
        \dfrac{r\kappa_{i}^D}{\log 100}\log \bigg( 100 - \dfrac{99 t}{t_{\text{end}}+l_2} \bigg), & \text{for } t_{\text{end}} \leq t < t_{\text{end}} + l_2 \\
    \end{cases}
\end{equation}
which captures the idea that demand for on-site consumption recovers very slowly initially and accelerates towards normal levels near the end of lockdown relaxation. $t_{\text{start}}$ and $t_{\text{end}}$ denote the start and end dates of the lockdowns. $l_2$ governs the number of days needed to ease out shocks after lockdown, and is assumed to equal eight weeks. $r$, whose value is inferred during the model calibration (Section \ref{section:sensitivity_analysis}), is introduced to quantify the magnitude of the household demand and other demand shock between both lockdowns, which did not fully recover to prepandemic levels (Fig.~\ref{fig:timeline}).

\paragraph{Other demand shocks} Other demand compromises: 1) Government and non-profit organizations, 2) gross fixed capital formation (investments), 3) exports of goods (economic activities A-E, Table \ref{tab:NACE21}), and, 4) exports of services (economic activities F-T, Table \ref{tab:NACE21}). We assume the shock to government and non-profit demand, as well as the shock to the exports of services follows the same time course as the household demand shock $\kappa_i^D (t)$ (Eq.~\eqref{eq:EPNM_timecourse_shocks}), as we expect these changes in consumption are all similarly driven by fear of infection or prohibition of consumption. During the second quarter of 2020, Belgium faced an investment shock of 16.2~\% \citep{oecd2023} and a shock to exports of goods of 25.0~\%, both shocks quickly recovered by July 2020 \citep{nbb2023b}. We assume the shocks to exports of goods and investments start recovering immediately after the initial shock, following the time course of Eq.~\eqref{eq:EPNM_timecourse_shocks}. Our choices result in an aggregated demand shock of 17~\% (Fig.~\ref{fig:timeline}). Opposed, Pichler et al. \cite{pichler2022} assumed there were no pandemic-induced shocks to government and non-profit demand and a 15~\% shock to exports of both goods and services.

\paragraph{Government furloughing} During the pandemic, households may experience income loss and this will in turn influence consumption behavior. The federal government can influence the economic outcome by compensating a fraction of the income loss, $b$, in order to mitigate reductions in household consumption. During the entire pandemic, the Belgian government furloughed up to 70~\% of lost labor income, and hence $b(t) = 0.7$ for the complete duration of all simulations shown in this work \citep{thuy2020}.

\paragraph{Labor supply shocks} Governments may close down industries or impose work-from-home orders, resulting in a supply shock that reduces the available amount of labor, mathematically denoted as $\kappa_{i}^S$. To inform the labor supply shocks under both lockdowns we use the percentage of temporarily unemployed workers obtained from the survey by the ERMG \citep{ermg2021}. To inform the labor supply shock during the first lockdown, the surveys from April 6, 2020, and April 13, 2020 were averaged resulting in an aggregated labor supply shock of 25\%. To inform the labor supply shock during the second lockdown, the surveys from November 10, 2020 and December 8, 2020 were averaged resulting in an aggregated labor supply shock of 8~\% (Fig.~\ref{fig:timeline}). No labor supply shock is imposed between both lockdowns and during the lockdown light. After lockdown, the shocks are gradually eased from the model using a ramp function of length $l_2 = 56~\text{days}$. Opposed, Pichler et al. \cite{pichler2022} used an index based on the product of the ability to work from home and the essentiality of the economic activity coined by del Rio-Chanona et al. \cite{delriochanona2020} to inform the labor supply shock.\\

Our work thus updates the magnitudes of the other demand shock $\kappa_i^F(t)$ and labor supply shocks $\kappa_i^S(t)$ to fall in line with observations during the 2020-2021 COVID-19 pandemic, and exploits data gathered during the COVID-19 pandemic to validate the magnitudes of the household demand shocks $\kappa_i^D(t)$.

\subsubsection{Demand}\label{section:EPNM_demand}

\paragraph{Total demand} The total demand of industry $i$ at time $t$, denoted $d_{i}(t)$, is the sum of the demand from all its customers,
\begin{equation}
    d_{i}(t) = \sum_{j=1}^{N} O_{ij}^d(t) + c_{i}^d(t) + f_{i}^d(t),
\end{equation}
where $O_{ij}^d(t)$ is the intermediate demand from industry $i$ to industry $j$, $c_{i}^d(t)$ is the total demand from households and $f_{i}^d(t)$ denotes other demand from governments, non-profits, investments, and exports. The superscript $d$ refers to \textit{desired}, as customer demand may not be met under the imposed shocks.\\

\begin{table}[!bh]
    \footnotesize
    \centering
    \caption[Overview of shocks caused by the COVID-19 pandemic used in the production network model of Pichler et al. (2022).]{Overview of shocks caused by the COVID-19 pandemic used in the production network model of Pichler et al. \cite{pichler2022}.}
    {\renewcommand{\arraystretch}{0.7}
    \begin{tabular}{>{\raggedright}p{1.5cm}>{\raggedright\arraybackslash}p{2.0cm}p{1.5cm}p{1.0cm}>{\centering\arraybackslash}p{1.4cm}}
        \toprule
        \textbf{Activity} & \textbf{Labor supply} & \multicolumn{2}{l}{\textbf{Demand}} & \textbf{On-site}\\ 
        ~                & \textbf{Lockdown 1} & \textbf{Households} & \textbf{Other} & \textbf{consumption}\\ 
        \midrule
        A01 & 0.0 & 10.0 & 14.5 & 0 \\
        A02 & 85.0 & 10.0 & 16.6 & 0 \\
        A03 & 0.0 & 10.0 & 14.7 & 0 \\
        B05-09 & 35.3 & 10.0 & 14.8 & 0 \\
        C10-12 & 0.6 & 10.0 & 14.8 & 0 \\
        C13-15 & 37.1 & 10.0 & 14.9 & 0 \\
        C16 & 61.1 & 10.0 & 12.9 & 0 \\
        C17 & 7.5 & 10.0 & 14.7 & 0 \\
        C18 & 6.0 & 10.0 & 14.2 & 0 \\
        C19 & 18.3 & 10.0 & 14.8 & 0 \\
        C20 & 2.6 & 10.0 & 15.0 & 0 \\
        C21 & 1.1 & 10.0 & 13.8 & 0 \\
        C22 & 28.3 & 10.0 & 14.8 & 0 \\
        C23 & 50.3 & 10.0 & 14.6 & 0 \\
        C24 & 57.7 & 10.0 & 14.8 & 0 \\
        C25 & 54.8 & 10.0 & 14.8 & 0 \\
        C26 & 38.5 & 10.0 & 14.9 & 0 \\
        C27 & 33.3 & 10.0 & 14.8 & 0 \\
        C28 & 49.7 & 10.0 & 14.9 & 0 \\
        C29 & 22.6 & 10.0 & 14.9 & 0 \\
        C30 & 48.8 & 10.0 & 15.0 & 0 \\
        C31-32 & 36.6 & 10.0 & 14.7 & 0 \\
        C33 & 3.3 & 10.0 & 15.0 & 0 \\
        D35 & 0.0 & 0.0 & 14.4 & 0 \\
        E36 & 0.0 & 0.0 & 15.0 & 0 \\
        E37-39 & 0.0 & 0.0 & 11.0 & 0 \\
        F41-43 & 35.6 & 10.0 & 15.0 & 0 \\
        G45 & 31.6 & 10.0 & 14.9 & 1 \\
        G46 & 23.6 & 10.0 & 14.4 & 0 \\
        G47 & 30.5 & 10.0 & 0.0 & 1 \\
        H49 & 11.1 & 67.0 & 11.7 & 1 \\
        H50 & 12.4 & 67.0 & 15.0 & 1 \\
        H51 & 0.1 & 67.0 & 15.0 & 1 \\
        H52 & 0.5 & 67.0 & 9.4 & 1 \\
        H53 & 0.0 & 0.0 & 15.0 & 1 \\
        I55-56 & 60.8 & 80.0 & 15.0 & 1 \\
        J58 & 14.4 & 0.0 & 15.0 & 0 \\
        J59-60 & 32.8 & 0.0 & 10.1 & 0 \\
        J61 & 0.9 & 0.0 & 15.0 & 0 \\
        J62-63 & 0.2 & 0.0 & 15.0 & 0 \\
        K64 & 0.0 & 0.0 & 15.0 & 0 \\
        K65 & 0.0 & 0.0 & 15.0 & 0 \\
        K66 & 0.0 & 0.0 & 15.0 & 0 \\
        L68 & 15.4 & 0.0 & 11.9 & 1 \\
        M69-70 & 2.0 & 0.0 & 15.0 & 1 \\
        M71 & 0.0 & 0.0 & 15.0 & 0 \\
        M72 & 0.0 & 0.0 & 13.5 & 0 \\
        M73 & 22.5 & 0.0 & 15.0 & 0 \\
        M74-75 & 3.0 & 0.0 & 15.0 & 0 \\
        N77 & 34.9 & 80.0 & 15.0 & 1 \\
        N78 & 34.9 & 0.0 & 15.0 & 0 \\
        N79 & 34.9 & 80.0 & 15.0 & 1 \\
        N80-82 & 34.9 & 0.0 & 15.0 & 0 \\
        O84 & 1.1 & 0.0 & 0.5 & 1 \\
        P85 & 0.0 & 0.0 & 0.1 & 1 \\
        Q86 & 0.1 & 0.0 & 0.0 & 1 \\
        Q87-88 & 0.1 & 0.0 & 0.0 & 1 \\
        R90-92 & 34.5 & 5.0 & 4.6 & 1 \\
        R93 & 34.5 & 2.0 & 1.8 & 1 \\
        S94 & 34.5 & 5.0 & 1.3 & 1 \\
        S95 & 34.5 & 5.0 & 15.0 & 1 \\
        S96 & 34.5 & 5.0 & 15.0 & 1 \\
        T97-98 & 0.0 & 0.0 & 0.0 & 1 \\
        \bottomrule
    \end{tabular}
    }
    \label{tab:overview_shocks_pichler}
\end{table}

\paragraph{Household demand} The household demand for good $i$ is,
\begin{equation}
    c_{i}^d(t) = \theta_{i}(t) \tilde{c}^d(t),
\end{equation}
where $\theta_{i}(t)$ is the household preference coefficient, denoting the share of good $i$ in the aggregate household demand $\tilde{c}^d(t)$. Before the pandemic, the share of good $i$ in total household consumption can be computed using the available data $\theta_{i}(0) = \textstyle c_{i}(0) / \sum_{j} c_{j}(0)$ (Table \ref{tab:overview_initial_states}). As household demand for good $i$ changes under the household demand shocks induced by the pandemic $\kappa_{i}^D(t)$, the consumption preference evolves dynamically according to,
\begin{equation}
    \theta_{i}(t) = \frac{\big(1-\kappa_{i}^D(t)\big)\theta_{i}(0)}{\textstyle \sum_j \big(1-\kappa_{j}^D(t)\big)\theta_{j}(0)}.
\end{equation}
The aggregate reduction in household demand caused by the pandemic shock is $\textstyle \tilde{\kappa}^D(t) = 1 - \sum_i \theta_{i}(0) (1-\kappa_{i}^D(t))$. However, households have the choice to save all the money they are not spending ($\Delta s = 1$), or to spend all their money on goods of other industries ($\Delta s = 0$). We can thus redefine the aggregate reduction in household demand shock as,
\begin{equation}
    \tilde{\kappa}^D(t) = \Delta s \left(1 - \sum_{i=1}^N \theta_{i}(0) \big(1-\kappa_{i}^D(t)\big)\right),
\end{equation}
where $\Delta s$ is the household savings rate. Aggregate household demand $\tilde{c}_t^d$ evolves according to following consumption function \citep{muellbauer2020,pichler2022},
\begin{multline}\label{eq:consumption}
    \tilde{c}^d(t) = \big(1 - \tilde{\kappa}^D(t) (1-\rho)\big) \\
    \exp \bigg( \rho \log \tilde{c}^d(t-1) + \frac{1-\rho}{2} \log \big(m \tilde{l}(t)\big) + \frac{1-\rho}{2} \log \big( m \tilde{l}^p(t) \big) \bigg),
\end{multline}
where $\tilde{\kappa}^D(t)$ is the aggregate reduction in household demand, $\rho$ is the adjustment speed of the aggregate household consumption, $m$ is the (pre-pandemic) share of labor income used by the households to consume goods, computed as $m = \sum c_i(0) / l_i(0) = 0.86$ for Belgium, $\tilde{l}(t)$ is the aggregated labor income and $\tilde{l}^p(t)$ is an estimation of permanent income. Nominally, the aggregated labor income would be computed as $\tilde{l}(t) = \textstyle \sum l_i(t)$. However, a government may choose to compensate a fraction $b(t)$ of the income losses. Hence, $\tilde{l}(t)$ in Eq.~\eqref{eq:consumption} is equal to,
\begin{equation}
    \tilde{l}(t) = \tilde{l}(0) + b(t) \big(\tilde{l}(0) - \tilde{l}(t)\big).
\end{equation}
Pessimistic expectations of permanent income may contribute to reduced demand \citep{muellbauer2020} and evolve dynamically during the pandemic. We assume income expectations are initially reduced to the labor income under the imposed labor supply shocks. Income expectations then gradually rise depending on the household's expectations of a quick V-shaped recovery versus a prolonged L-shaped recession. Mathematically this can be expressed as,
\begin{equation}
\tilde{l}^p(t) = 
\begin{cases}
\tilde{l}(0), & \text{ } t \leq \text{15-03-2020} \\
\big(1 - \rho + \rho \zeta(t-1) - (1-\rho)(1-\zeta_L)/L\big)\tilde{l}(0), & \text{ } t > \text{15-03-2020} \\
\end{cases}
\end{equation}
where $L$ denotes the fraction of households believing in an L-shaped recovery and $\zeta_L$ is the reduction in income at the start of the first lockdown,
\begin{equation}
\zeta_L = 1 - \dfrac{\tilde{l}(0) - \sum_i \kappa_{i}^S(t) l_{i}(0)}{\tilde{l}(0)}.
\end{equation}
Consumer confidence, as surveyed by the \cite{nbb2021}, followed a V-shaped recovery and hence $L=1$ (Fig.~\ref{fig:permanent_income}).

\paragraph{Other demand} Industry $i$ faces demand $f_{i}^d(t)$ from sources not explicitly included in the model. These sources are: 1) Government and non-profit organizations, 2) gross fixed capital formation (investments), 3) exports of goods (economic activities A-E, Table \ref{tab:NACE21}), and, 4) exports of services (economic activities F-T, Table \ref{tab:NACE21}). During the 2020-2021 COVID-19 pandemic, the other demand is scaled with the shock $\kappa_i^F(t)$ discussed prior (Fig.~\ref{fig:timeline}). Mathematically,
\begin{equation}
f_i^d(t) = \left( 1 - \kappa_i^F(t) \right) f_i(0).
\end{equation}

\paragraph{Intermediate demand} For industry $j$ to produce one unit of output, inputs from industry $i$ are needed. The production recipe is encoded in the matrix of technical coefficients $\mathcal{A}$, where an element $\mathcal{A}_{ij} = Z_{ij}/x_{j}(0)$ represents the expense in inputs $i$ to produce one unit of output $j$. In the model, each industry $j$ aims to keep a target inventory of inputs $i$, $n_j Z_{ij}(0)$, so that production can go on for $n_j$ more days (Table \ref{tab:overview_initial_states}). The stock of inputs $i$ kept by industry $j$ is denoted as $S_{ij}(t)$. The intermediate demand faced by industry $j$ from industry $i$ at time $t$ is modeled as the sum of two components,
\begin{equation}
O_{ij}^d(t) = \mathcal{A}_{ij}d_{j}(t-1) + \frac{1}{\tau} \big( n_j Z_{ij}(0) - S_{ij}(t-1) \big).
\end{equation}
The first term represents the attempts of industry $j$ to satisfy incoming demand under the naive assumption that demand on the day $t$ will be the same as on day $t-1$. The second term represents the attempts to close inventory gaps. The parameter $\tau$ governs how quickly an industry aims to close inventory gaps and ranges from 1  to 30 days have been proposed \citep{henriet2012, hallegatte2012}. We determine its optimal value during the calibration (Section \ref{section:sensitivity_analysis}).

\subsubsection{Supply}\label{section:EPNM_supply} Every industry aims to satisfy the incoming demand by producing the required output. However, production under the imposed pandemic shocks is subject to two constraints.

\paragraph{Labor supply constraints} Productive capacity is assumed to linearly depend on the available amount of labor and hence,
\begin{equation}\label{eq:EPNM_labor_supply_constraints_1}
    x_{i}^{\text{cap}}(t) = \frac{l_{i}(t)}{l_{i}(0)} x_{i}(0).
\end{equation}
Recall that during the lockdowns, labor supply is shocked and the maximum amount of available labor is reduced to,
\begin{equation}\label{eq:EPNM_labor_supply_constraints_2}
    l_{i}^{max}(t) = \big(1 - \kappa_{i}^S(t)\big) l_{i}(0).
\end{equation}
However, as explained in Section \ref{section:EPNM_hiring_firing}, industries are allowed to fire workers if productive capacity is greater than demand, and thus, the output can be constrained further by a shortage of labor. By combining Eq.~\eqref{eq:EPNM_labor_supply_constraints_1} and Eq.~\eqref{eq:EPNM_labor_supply_constraints_2}, 
\begin{equation}
    x_{i}^{\text{cap}}(t) \leq \big(1 - \kappa_{i}^S(t)\big) x_{i}(0).
\end{equation}

\paragraph{Input bottlenecks} The productive capacity of an industry can be constrained if an insufficient supply of inputs is in stock, referred to as a ``production function" ($\mathcal{F}$). In a classical Leontief approach, every input encoded in the recipe matrix $\mathcal{A}_{ij}$ is considered critical to production. By using a survey on the criticality of inputs to different industries, the Leontief production function can be relaxed to a partially-binding (PBL) Leontief production function \citep{pichler2022}. The inputs $j$ of industry $i$ are either important or critical to production, denoted $\mathcal{I}_{ij}$ and $\mathcal{C}_{ij}$ respectively. Input bottlenecks can be treated in the following ways, ranked from most to least restrictive.
\begin{enumerate}
    \item Leontief: Every input encoded in the matrix of technical coefficients $\mathcal{A}_{ij}$ is binding, i.e. the depletion of one input, regardless of its actual relevance to production, halts production. Mathematically,
        \begin{equation}
            x_{i}^{\text{inp}}(t) =  \min_{j} \bigg\{ \frac{S_{ji}(t)}{\mathcal{A}_{ji}} \bigg\}.
        \end{equation}
    \item Strongly critical: Inputs rated \textit{critical} and \textit{important} are binding. Thus, if an input rated \textit{critical} or \textit{important} is depleted in an industry's stock, production is halted. Mathematically,
        \begin{equation}
            x_{i}^{\text{inp}}(t) =  \min_{j \in \{ \mathcal{C}_i \cup \mathcal{I}_i \}} \bigg\{ \frac{S_{ji}(t)}{\mathcal{A}_{ji}} \bigg\}.
        \end{equation}
    \item Half-critical: An intermediate case in which the depletion of \textit{critical} inputs halts production completely while the depleting \textit{important} inputs reduce production with 50~\%. Mathematically,
        \begin{equation}
            x_{i}^{\text{inp}}(t) = \min_{ \{j \in \mathcal{C}_i, k \in \mathcal{I}_i\}} \bigg\{ \frac{S_{ji}(t)}{\mathcal{A}_{ji}}, \frac{1}{2} \bigg( \frac{S_{ki}(t)}{\mathcal{A}_{ki}} + x_{i}^{\text{cap}}(0) \bigg)  \bigg\}.
        \end{equation}
    \item Weakly critical: All \textit{important} inputs are treated as \textit{non-critical} inputs and thus do not influence productive capacity.
        \begin{equation}
            x_{i}^{\text{inp}}(t) =  \min_{j \in \mathcal{C}_i} \bigg\{ \frac{S_{ji}(t)}{\mathcal{A}_{ji}} \bigg\}.
        \end{equation}
    \item Linear: All inputs are perfect substitutes, production can go on as long as there are other inputs.
        \begin{equation}
            x_{i}^{\text{inp}}(t) = \frac{\sum_j S_{ji}(t)}{\sum_j \mathcal{A}_{ji}}
        \end{equation}
\end{enumerate}
Both the Leontief production function and the linear production function are unrealistic production functions. In this work, we will use the available data to identify the most appropriate PBL production function (Section \ref{section:sensitivity_analysis}). A more strict PBL production function will result in larger supply chain bottlenecks and will thus lower the predicted output. However, the restocking rate $\tau$ introduced prior has a similar influence on the predicted output. Closing inventory gaps slowly results in larger supply chain bottlenecks under pandemic shocks. Certain combinations of PBL production functions and restocking rates can thus lead to similar model projections (Fig.~\ref{fig:sensitivity-tau}).

\subsubsection{Realised output and rationing}\label{section:EPNM_rationing} As each industry aims to maximally satisfy incoming demand under its production constraints, the realized output of sector $i$ at time $t$ is,
\begin{equation}\label{eq:realised_output}
    x_{i} (t) = \min \{ x_{i}^{\text{cap}}(t),\ x_{i}^{\text{inp}}(t),\ d_{i}(t)\},
\end{equation}
thus, the output is constrained by the smallest of three values: the labor-constrained productive capacity $x_{i}^{\text{cap}}(t)$, the input-constrained productive capacity $x_{i}^{\text{inp}}(t)$ and total demand $d_{i}(t)$. If productive capacity was lower than total demand, industries ration their output equally across customers (\textit{strict proportional rationing}), mathematically,
\begin{equation}
    c_{i}(t) =  c_{i}^d(t) \bigg( \frac{x_{i}(t)}{d_{i}(t)} \bigg),
\end{equation}
\begin{equation}
    f_{i}(t) = f_{i}^d(t) \bigg( \frac{x_{i}(t)}{d_{i}(t)} \bigg),
\end{equation}
\begin{equation}
    O_{ij}(t) = O_{ij}^d(t) \bigg( \frac{x_{i}(t)}{d_{i}(t)} \bigg).
\end{equation}
The performance of alternative rationing schemes proposed by Pichler et al. \cite{pichler2021} were assessed but their performance deemed unsatisfactory.

\subsubsection{Inventory adjustment}\label{section:EPNM_inventory_updating}
After the realized output has been rationed among the customers, inventories can be updated,
\begin{equation}
    S_{ij}(t) = \max_{i,j} \{ S_{ij}(t-1) + O_{ij}(t) - \mathcal{A}_{ij} x_{j}(t),\ 0\},
\end{equation}
the new stocks of input $i$ in industry $j$ is thus equal to the intermediate inputs $i$ received minus the inputs $i$ consumed in the production of $j$ outputs.

\subsubsection{Hiring and firing}\label{section:EPNM_hiring_firing}
Firms will adjust their labor force depending on what constraint was binding in Eq.~\eqref{eq:realised_output}. If the supply of labor, $x_{i}^{\text{cap}}(t)$, was binding then industry $i$ will attempt to hire as many workers as needed to make the supply of labor not binding. Opposed, if either input constraints $x_{i}^{\text{inp}}(t)$ or total demand $d_{i}(t)$ were binding, industry $i$ will attempt to lay off workers until labor supply constraints become binding,
\begin{equation}
    \Delta l_{i}(t) = \frac{l_{i}(0)}{x_{i}(0)} \big[ \min \{ x_{i}^{\text{inp}}(t), d_{i}(t) \} - x_{i}^{\text{cap}(t)} \big].
\end{equation}
However, the process of adjusting the labor force is not an instantaneous one and takes time. Thus,
\begin{equation}
    l_{i}(t) =
    \begin{cases}    
        l_{i}(t-1) + 1/\iota_H \Delta l_{i}(t), & \text{if } \Delta l_{i}(t) \geq 0, \\
        l_{i}(t-1) + 1/\iota_F \Delta l_{i}(t), & \text{if } \Delta l_{i}(t) < 0, \\
    \end{cases}
\end{equation}
where $\iota_H$ is the average time needed to hire a new employee and $\iota_F$ is the average time needed to lay off an employee. Similar to \cite{pichler2022}, we assume that hiring takes twice as long as firing. We assess the influence of $\iota_F$ and $\iota_H$ on the predictive accuracy of the model in the sensitivity analysis (Section \ref{section:sensitivity_analysis}). We assume that in the Public Administration (O84) and Education (P85), no firing takes place during the pandemic.

\subsection{Calibration and sensitivity analysis}\label{section:sensitivity_analysis}

\subsubsection{Available data and objective function}
Four time series of economic data were retrieved: 1) The number of B2B transactions were retrieved from a bank with a market share of $\geq 25~\%$ in Belgium, 2) the synthetic GDP was retrieved from the Belgian National Bank \citep{NBBstat}, 3-4) the revenue and employment surveys were conducted by the ERMG \cite{ermg2021}. These time series of data are characterized by a temporal axis and a sectoral axis, i.e. data on these four economic indicators is available at different dates and for different economic activities included in the NACE classification (Table~\ref{tab:time_series}, Table~\ref{tab:EPNM_overview_data}). In total, 115 time series at the level of economic activities and three aggregated (national) time series are available. The data are normalized with their pre-pandemic values so that all time series are expressed as a percentage reduction compared to pre-pandemic levels.\\

\begin{table}[h!]
    \caption[Characteristics of the economic time series for Belgium.]{Characteristics of the economic time series for Belgium: The model state used as a proxy for the economic indicator, the total number of temporal data available from 31-03-2020 until 31-03-2021, the sectoral aggregation of the available time series, where BE refers to the national aggregate. The total number of time series available is included between parenthesis.}
    \centering
    {\footnotesize\renewcommand{\arraystretch}{1.10}
    \begin{tabular}{p{2.5cm}p{1.2cm}p{1.2cm}p{2.5cm}p{2.5cm}}
        \toprule
        \textbf{Indicator} & \textbf{Corr. Model state} & \textbf{No. timesteps} & \textbf{Aggregation} &  \textbf{Source} \\
        \midrule
        B2B transactions & $\sum_j O_{ij}(t)$ & 60 & NACE 21 (20) & Belgian bank (market share $\geq 25~\%$) \\
        Synthetic GDP & $x_{i}(t)$ & 14 & BE + NACE 64 (21) & Business surveys, NBB \cite{NBBstat} \\ 
        Revenue survey & $x_{i}(t)$ & 18 & BE + NACE 64 (37) & ERMG \cite{ermg2021} \\ 
        Employment survey & $l_{i}(t)$ & 17 & BE + NACE 64 (37) & ERMG \cite{ermg2021} \\ 
        \bottomrule
    \end{tabular}
    }
    \label{tab:time_series}
\end{table}

To assess the model's performance in matching these time series, the value-weighted mean absolute error $\text{MAE}_{\text{vw}}$ is used as \textit{objective function}. For the synthetic GDP, proxied by gross output $x_i(t)$, 
\begin{equation}
    \text{MAE}_{\text{vw}, \text{GDP}}(t) = \sum_{i\in \mathcal{D}} \bigg( \dfrac{x_i(0)}{\sum_i x_i(0)} \bigg) \|x_i(t) - \hat{x}_i(t)\|,
\end{equation}
where $x_i(0)$ is the gross output of economic activity $i$ prior to the pandemic (Table \ref{tab:overview_initial_states}). $x_i(t)$ is the reported gross output of economic activity $i$ at time $t$ and $\hat{x}_i(t)$ is the corresponding model prediction. Both $x_i(t)$ and $\hat{x}_i(t)$ are expressed in percentage reduction in GDP compared to pre-pandemic levels. $\mathcal{D}$ is the collection of economic activities $i$ for which a time series of synthetic GDP is available (Table \ref{tab:EPNM_overview_data}). The total $\text{MAE}_{\text{vw}}$ is computed as the mean over all available economic indicators and timesteps. The $\text{MAE}_{\text{vw}}$ was chosen over the weighted sum of squared errors (WSSE)\footnote{Where the weight is chosen as the inverse of the variance, which is assumed to be proportional to the magnitude of the measurements \cite{motulsky2005}. So, $\text{WSSE} = w_i(t) \sum_{i \in \mathcal{D}} (x_i(t) - \hat{x}_i(t))^2$, where $w_i(t) = 1/\sigma^2_i(t) \propto 1/x_i^2(t)$.} and mean average percentage error (MAPE), as these indicators assign a disproportionate weight to observations close to zero. As a measures for bias on the model's projections, we use the value-weighted mean error ($\text{ME}_{\text{vw}}$), which is computed as follows for the synthetic GDP,
\begin{equation}\label{eq:ME_vw}
    \text{ME}_{\text{vw}, \text{GDP}}(t) = \sum_{i\in \mathcal{D}} \bigg( \dfrac{x_i(0)}{\sum_i x_i(0)} \bigg) \left( x_i(t) - \hat{x}_i(t)\right).
\end{equation}

\subsubsection{Parameters}

To identify their optimal value and sensitivity, we compute the total $\text{MAE}_{\text{vw}}$ on a grid spanning eight parameters of interest: the production function $\mathcal{F}$, the household consumption shock to Manufacturing and Construction (C, F), the household consumption shock to Retail (G46, G47, S95), the household consumption shock to land, sea and air transport (H49, H50, H51), and, the household consumption shock to strongly consumer-facing activities (I55-56, N77, N79, R90-92, R93, S94, S96), the fraction of the household and other demand shocks remaining between both lockdowns and during the lockdown light $r$, the speed of inventory restocking $\tau$, and, the speed of hiring and firing, $\iota_F$ and $\iota_H$ (Table \ref{tab:sensitivity_overview_parameters}). The grid spans \num{468 750} combinations, resulting in a computation of approximately eight days on an Intel Xeon W-2295 CPU (18C @ 3.00GHz).\\

The household consumption shock to the wholesale of motor vehicles (G45) was set to 90~\% as this resulted in the best match between the model and the empirical data (Table \ref{tab:overview_shocks}). The aggregate household consumption adjustment speed ($\rho$), the changes in the savings rate ($\Delta S$), the fraction of households believing in a V-shaped economic recovery ($L$), and the reimbursed fraction of lost labor income ($b$) were also excluded from the sensitivity analysis because approximate estimates for these parameters are available and the sensitivity of the model output to changes in the values of these parameters is low (Fig.~\ref{fig:sensitivity_unimportant_parameters}). The labor supply shock $\kappa_{i}^S$ is informed using data by the ERMG \cite{ermg2021}, however, the model output is very sensitive to changes in the labor supply shock (Fig.~\ref{fig:sensitivity_unimportant_parameters}).

\begin{table}[!h]
    \centering
    \caption{Overview of model parameters and their values used in the sensitivity analysis.}
    {\footnotesize\renewcommand{\arraystretch}{1.10}
    \begin{tabular}{p{0.6cm}p{5.5cm}>{\raggedright\arraybackslash}p{4.5cm}}
        \toprule
        \textbf{Symb.} & \textbf{Name} & \textbf{Values} \\
        \midrule
        $\mathcal{F}$ & Production function & Leontief, Strongly critical, Half-critical, Weakly critical, Linear \\
        $\kappa_{i}^D$ & Household demand shock: Manufacturing and Construction (C, F) & $0, 15, 30, 45, 60~(\%)$ \\
        $\kappa_{i}^D$ & Household demand shock: Retail (G46, G47) & $0, 10, 20, 30, 40~(\%)$ \\
        $\kappa_{i}^D$ & Household demand shock: Land, sea and air transport (H49, H50, H51) & $0, 20, 40, 60, 80, 100~(\%)$ \\
        $\kappa_{i}^D$ & Household demand shock: Consumer-facing activities (I55-56, N77, N79, R90-92, R93, S94, S96) & $80, 85, 90, 95, 100~(\%)$ \\
        $r$ & Fraction of household and other demand shock under lockdown, remaining between both lockdowns and during the \textit{lockdown light} & $0, 25, 50, 75, 100~(\%)$ \\
        $\tau$ & Speed of inventory restocking & $1, 7, 14, 21, 28~(\text{days})$ \\
        $\iota_F$ & Speed of firing & $1, 7, 14, 21, 28~(\text{days})$ \\
        $\iota_H$ & Speed of hiring & $2\iota_F$, assumed 
        \citep{pichler2022} \\
      \bottomrule
    \end{tabular}
    }
    \label{tab:sensitivity_overview_parameters}
\end{table}

\section{Results and discussion}\label{section:EPNM_results}

\begin{figure}[h!]
    \centering
    \includegraphics[width=1.02\linewidth]{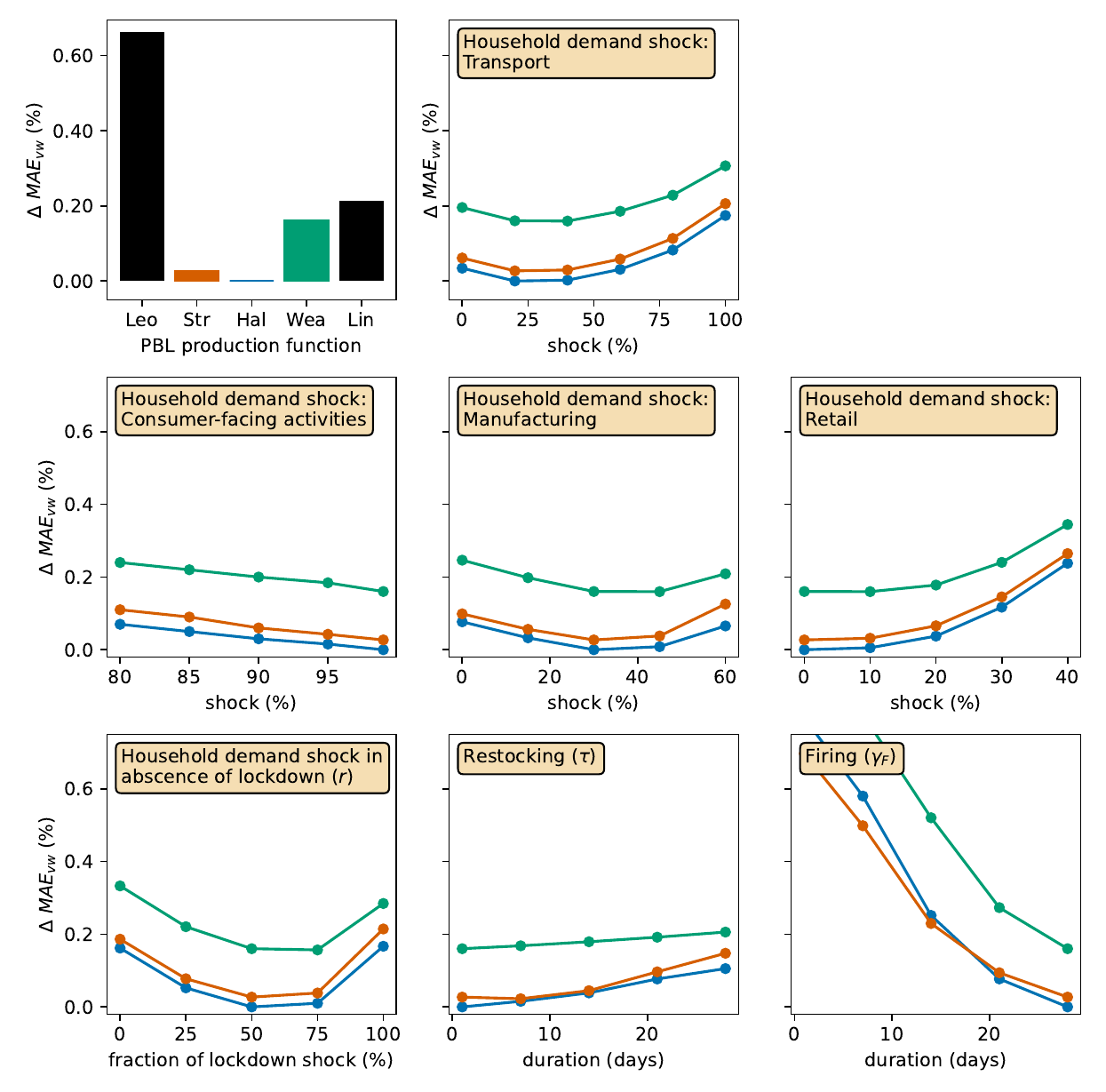}
    \caption[Sensitivity of the value-weighted mean absolute error to changes in the production function, household consumption shocks, restocking rate, and firing rate.]{Sensitivity of the $\text{MAE}_{\text{vw}}$ (as compared to the optimum, $\Delta \text{MAE}_{\text{vw}}$) to changes in the production function, household consumption shocks, restocking rate, and firing rate. The presented results are the one-dimensional slices through the global minimum obtained in the grid search.} 
    \label{fig:sensitivity_slice_minimum}
\end{figure}

The minimal $\text{MAE}_{\text{vw}}$ on the grid spanning the parameter values in Table \ref{tab:sensitivity_overview_parameters} was $4.71~\%$, implying that on average, model projections, over all quarters, sectors and economic indicators deviate 4.71~\% from the observations. The $\text{ME}_{\text{vw}}$  was equal to $-0.45~\%$, implying the model slightly overestimates the economic damages (Eq.~\eqref{eq:ME_vw}). The production function resulting in the lowest $\text{MAE}_{\text{vw}}$ is the half-critical PBL production function (Table \ref{tab:sensitivity_global_minimum}). The Leontief production function is the most unfavorable production function, followed by the linear and weakly critical production functions (Fig.~\ref{fig:sensitivity_slice_minimum}). The difference in accuracy between the half-critical and strongly critical production functions is very small (0.03~\%), so the data do not have sufficient power to distinguish between these two production functions  (Figs. \ref{fig:sensitivity_slice_minimum} and \ref{fig:sensitivity_prodfunc_tau}). Discriminatory power would have been greater if the lockdown had lasted longer, as secondary shocks caused by supply-chain disruptions become more pronounced over time (Fig.~\ref{fig:sensitivity-tau}). Relaxing the Leontief production function to the PBL production functions clearly results in a significant improvement of the model's accuracy in light of empirical data (Fig.~\ref{fig:sensitivity_slice_minimum}).\\

The optimal household consumption shocks that resulted in the best agreement between the model and the empirical data differed from those proposed by the US Congressional Budget Office \cite{congressional_budget_office2006} and used by Pichler et al. \cite{pichler2022}. Still, the impact of these changes in shocks to the model's accuracy is limited (Table \ref{tab:sensitivity_overview_parameters}). The model is not very sensitive to changes in the household consumption shock to manufacturing and construction, transport, and consumer-facing activities. The model is sensitive to changes in the household demand shock to retail, most likely because retail is highly connected to other economic activities in the Belgian production network (Fig.\ref{fig:sensitivity_slice_minimum}). The model is not sensitive to the restocking rate ($\tau$), although lower restocking rates result in a slightly increased model accuracy. Lastly, the model is most sensitive to the choice of firing rate ($\iota_F$), with a clear minimum in $\text{MAE}_{\text{vw}}$ at $\iota_F = 28~d$. However, an average of four weeks to lay off employees seems high, especially in light of the furloughing set up by the Belgian governments, which made temporarily laying off employees easier.\\

\begin{table}[hb!]
    \centering
    \caption[Optimal parameter values found in the production network model's calibration.]{Parameter values resulting in the minimal total $\text{MAE}_{\text{vw}} = 4.71~\%$ (``optimal" parameters).}
    {\footnotesize\renewcommand{\arraystretch}{1.10}
    \begin{tabular}{p{0.6cm}p{4.0cm}>{\centering}p{1.8cm}>{\centering}p{1.5cm}>{\centering\arraybackslash}p{1.5cm}}
        \toprule
        \textbf{Symb.} & \textbf{Name} & \textbf{Optimal value} & \textbf{Pichler et al. \citep{pichler2022}} & $\bm{\Delta \text{\textbf{MAE}}_{\text{\textbf{vw}}}}$\\ \midrule
        $\mathcal{F}$ & Production function & Half-critical & Half-critical & 0.0~\%\\
        $\kappa_{i}^D$ & Household demand shock: Manufacturing and Construction (C, F) & 30~\% & $10~\%$ & 0.05~\% \\
        $\kappa_{i}^D$ & Household demand shock: Retail (G46, G47) & 0~\% & 10~\% & 0.01~\% \\
        $\kappa_{i}^D$ & Household demand shock: Land, sea and air transport (H49, H50, H51) & 30~\% & 67~\% & 0.05~\% \\
        $\kappa_{i}^D$ & Household demand shock: Consumer-facing activities (I55-56, N77, N79, R90-92, R93, S94, S96) & 100~\% & 80~\% & 0.07~\% \\
        $r$ & Fraction of household and other demand shock under lockdown, remaining between both lockdowns and during the \textit{lockdown light} & 50~\% & NA & NA \\
        $\tau$ & Speed of inventory restocking & 1~day & 10~days & 0.02~\% \\
        $\iota_F$ & Speed of firing & 28~days & 14~days & 0.25\% \\
      \bottomrule
    \end{tabular}
    }
    \label{tab:sensitivity_global_minimum}
\end{table}

\begin{figure}[th!]
    \centering
    \includegraphics[width=\linewidth]{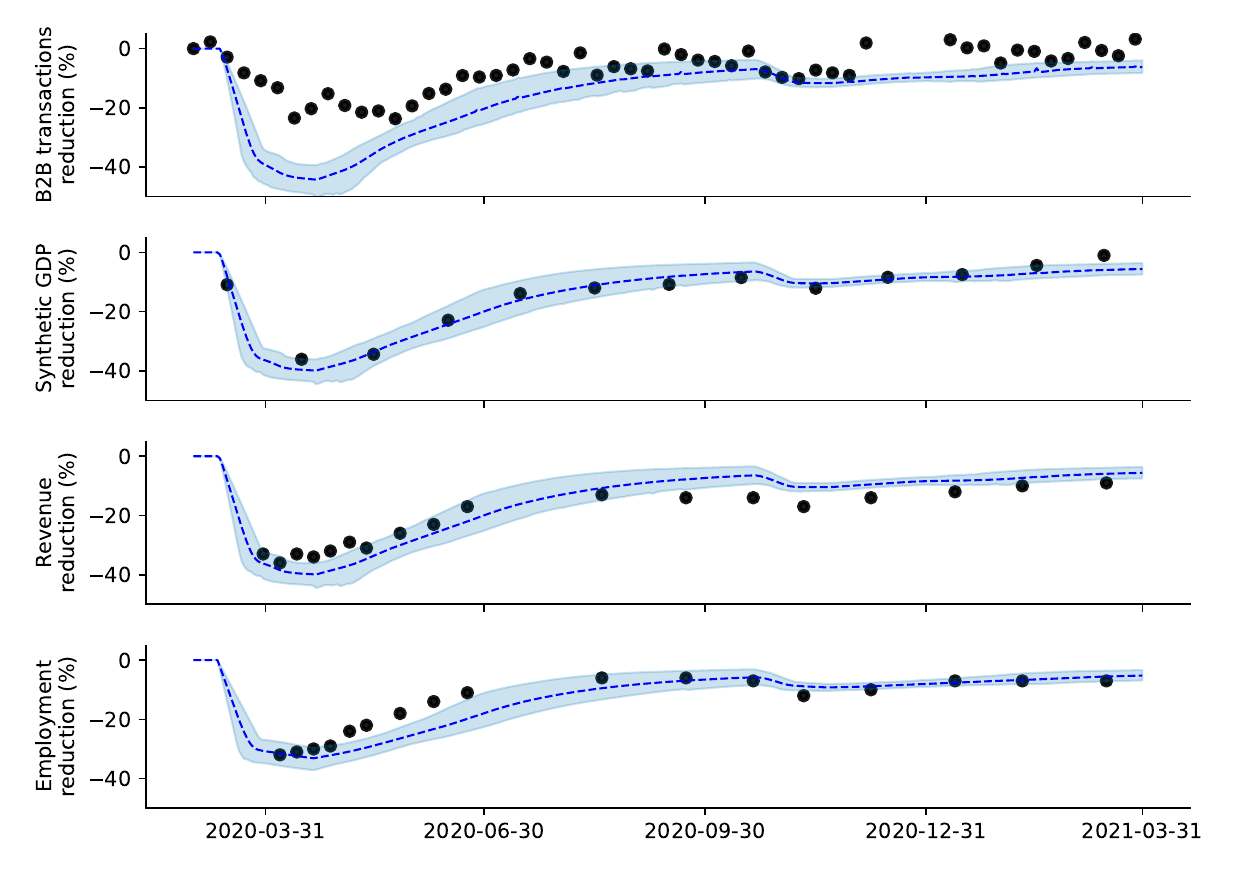}
    \caption[Comparison between the production network model projections and available data (aggregated).]{Comparison between the model projections and available data (aggregated). Using the optimal parameter values found during the sensitivity analysis (Table \ref{tab:sensitivity_global_minimum}). From top to bottom: B2B transactions, Synthetic GDP, Revenue survey, and employment survey.} 
    \label{fig:plot-fit-national}
\end{figure}

The model consistently provides accurate aggregated projections for GDP, revenue, and employment throughout the entire pandemic. However, there is a notable overestimation of the reduction in B2B transactions during the initial lockdown, pointing to epistemic uncertainty in the model. In Fig.~\ref{fig:plot-fit-national}, the modeled reduction in B2B transactions at the start of the pandemic is more abrupt than the observed reduction, suggesting there were incentives for businesses to continue trading goods and services working against the pandemic's induced supply and demand shocks. In the model, the pandemic's induced shocks immediately constrain the modeled number of B2B transactions. In reality, businesses may possess a backlog of outstanding orders and contractual obligations to deliver these goods and services despite the pandemic's disruptions. To enhance the model's realism, we could consider incorporating the notion that each sector maintains an inventory of finished products awaiting delivery, in a fashion similar to the inventory of inputs. Additionally, deferral of payments \citep{nbb2020} which is not included in the model may have further smoothed the shock to B2B transactions. However, including the impact of fiscal policies on business liquidity would drastically increase the model's complexity.\\

In Figure \ref{fig:plot-fit-sectors}, the model projections and the available data can be compared at the sectoral level and across four quarters 2020Q2, 2020Q3, 2020Q4, and 2021Q1 (columns) and four economic indicators (rows). In each panel, the $\text{MAE}_{\text{vw}}$ and $\text{ME}_{\text{vw}}$ are given to asses the average accuracy and bias of the model's output. Consistent with the aggregated projections, the model's projections of GDP, revenue, and employment are accurate. The accuracy of the reduction in employment is especially high (2.0~\%) and unbiased (0.2~\%). In contrast, the reduction in B2B demand during the first COVID-19 lockdown (2020Q2) is not very accurate (12.8~\%) and has a strong pessimistic bias (-11.5~\%).\\

The largest shocks during the COVID-19 pandemic were observed in the accommodation (I), rental \& leasing (N77), travel (N79), recreation (R), and personal services (S), which is not surprising considering these sectors experience the largest shocks in demand, both from households and from government and non-profits (Fig.~\ref{fig:plot-fit-sectors}). The simulated reductions for travel agencies (N79) in between the first and second lockdowns (2020Q3) are too optimistic, indicating demand for these sectors had not properly recovered during the summer of 2020. We conclude that using the optimal parameters, the model provides accurate projections of GDP, revenue, and employment reductions, but not of the reduction in B2B transactions.\\

\begin{figure}[th!]
    \centering
    \includegraphics[width=0.85\linewidth]{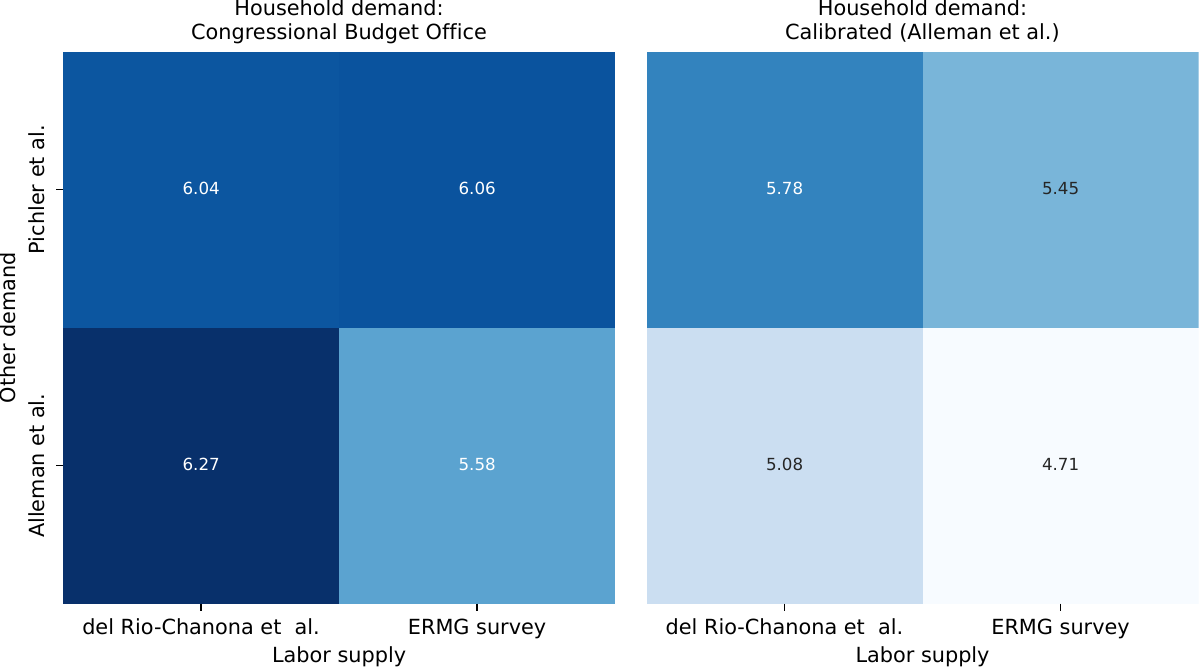}
    \caption{Effect of altering the labor supply shock, household demand shock, and other demand shock on the total (value-weighted) mean absolute error.}
    \label{fig:comparing_shocks}
\end{figure}

Lastly, our model differs from Pichler et al. \cite{pichler2022} in three ways: 1) We used survey data by the ERMG on the number of temporarily unemployed workers during the COVID-19 pandemic to inform the magnitude of the labor supply shock \citep{ermg2021}. 2) We refined the shocks to demand by governments and non-profits, investements, and exports to fall in line with observed trade data. 3) We calibrated the household demand shock under lockdown using 115 time series of aggregated and sectoral data. By incorporating these changes, our model achieves a higher overall accuracy of 4.71~\% as opposed to 6.04~\% for the model and shocks of Pichler et al. \cite{pichler2022} (Fig.~\ref{fig:comparing_shocks}, Fig.~\ref{fig:EPNM_comparison}). Using the calibrated household demand shocks (Fig.~\ref{fig:comparing_shocks}, left) instead of the estimates of the US Congressional Budget Office \cite{congressional_budget_office2006}  (Fig.~\ref{fig:comparing_shocks}, right) improves the overall accuracy. Similarly, using the labor supply shock informed by the surveys of the ERMG \cite{ermg2021} instead of the index of del Rio-Chanona et al. \cite{delriochanona2020} also improved the model's accuracy. Using the household demands shocks proposed by the US Congressional Budget Office\cite{congressional_budget_office2006}, the labor supply shock by del Rio-Chanona et al. \cite{delriochanona2020} and the other demand shocks used in this work lowers the accuracy to 6.27~\%. Our three modifications to the model of \cite{pichler2022} thus work synergistically to improve accuracy in light of the empirical data (Fig.~\ref{fig:comparing_shocks}).

\clearpage

\section{Conclusions}

In this study, we implemented the dynamic production network model proposed at the start of the COVID-19 pandemic by Pichler et al. \cite{pichler2022}. We refined the propagated shocks to align with observed data collected during the pandemic and calibrated some less-informed parameters and consumption shocks using 115 economic time series. Our findings indicate that the refined model is more effectively able to capture the evolution of GDP, revenue, and employment in Belgium during the pandemic at both individual economic activity and aggregate levels than the original model. However, the model struggles to accurately capture the reduction in B2B demand during the pandemic, as evidenced by its overestimation of the decline in B2B transactions, indicating potential shortcomings in accounting for businesses' incentives to maintain trade despite induced supply and demand shocks. We confirm the relaxation of the stringent Leontief production function by means of a survey on the criticality of inputs significantly improved the model's accuracy \citep{pichler2022}. However, in spite of the large amount of data used in this work, distinguishing between varying degrees of relaxation in the production function proved challenging. Additionally, we found that utilizing business surveys on pandemic-induced unemployment to inform the labor supply shock significantly improved the model's precision. Overall, our study demonstrates the model's validity in assessing the impact of economic shocks caused by an epidemic in Belgium.\\

%%%%%%%%%%%%%%%%%%%%%%%%
%% Bureaucratic stuff %%
%%%%%%%%%%%%%%%%%%%%%%%%

\clearpage

\bmhead{Supplementary information}

This work contains an index of the NACE Rev. 2 classification in 63 economic activities, and 21 economic activities (Tables \ref{tab:NACE21} and \ref{tab:NACE64}), the initial model states and the targeted inventory (Table \ref{tab:overview_initial_states}), as well as several simulations in support of the main text (Fig.~\ref{fig:impact_discrete_continuous}, \ref{fig:permanent_income}, \ref{fig:sensitivity_unimportant_parameters}, \ref{fig:sensitivity_prodfunc_tau}), and a sectoral breakdown of the available economic data used in this work (Table \ref{tab:EPNM_overview_data}).\\

\bmhead{Author contributions}

\textbf{Tijs W. Alleman}: Conceptualisation, Software, Methodology, Investigation, Writing – original draft. \textbf{Koen Schoors} Data acquisition, Conceptualisation. \textbf{Jan M. Baetens}: Investigation, Funding acquisition, Supervision, Project administration, Writing – original draft.\\

\bmhead{Acknowledgements}

This work was financially supported by \textit{Crelan}, the \textit{Ghent University Special Research Fund} (BOF), by the \textit{Research Foundation Flanders} (FWO), Belgium, project numbers G0G2920N and 3G0G9820, and, by \textit{VZW 100 km Dodentocht Kadee}, through the organization of the 2020 100 km COVID-Challenge.\\

\bmhead{Conflict of interest}
The authors declare that they have no known competing financial interests or personal relationships that could have appeared to influence the work reported in this paper. The funding sources played no role in study design; in the collection, analysis, and interpretation of data; in the writing of the report; and in the decision to submit the article for publication.\\

\begin{landscape}
\thispagestyle{empty}
\begin{figure}[h!]
    \centering
    \includegraphics[width=0.85\linewidth]{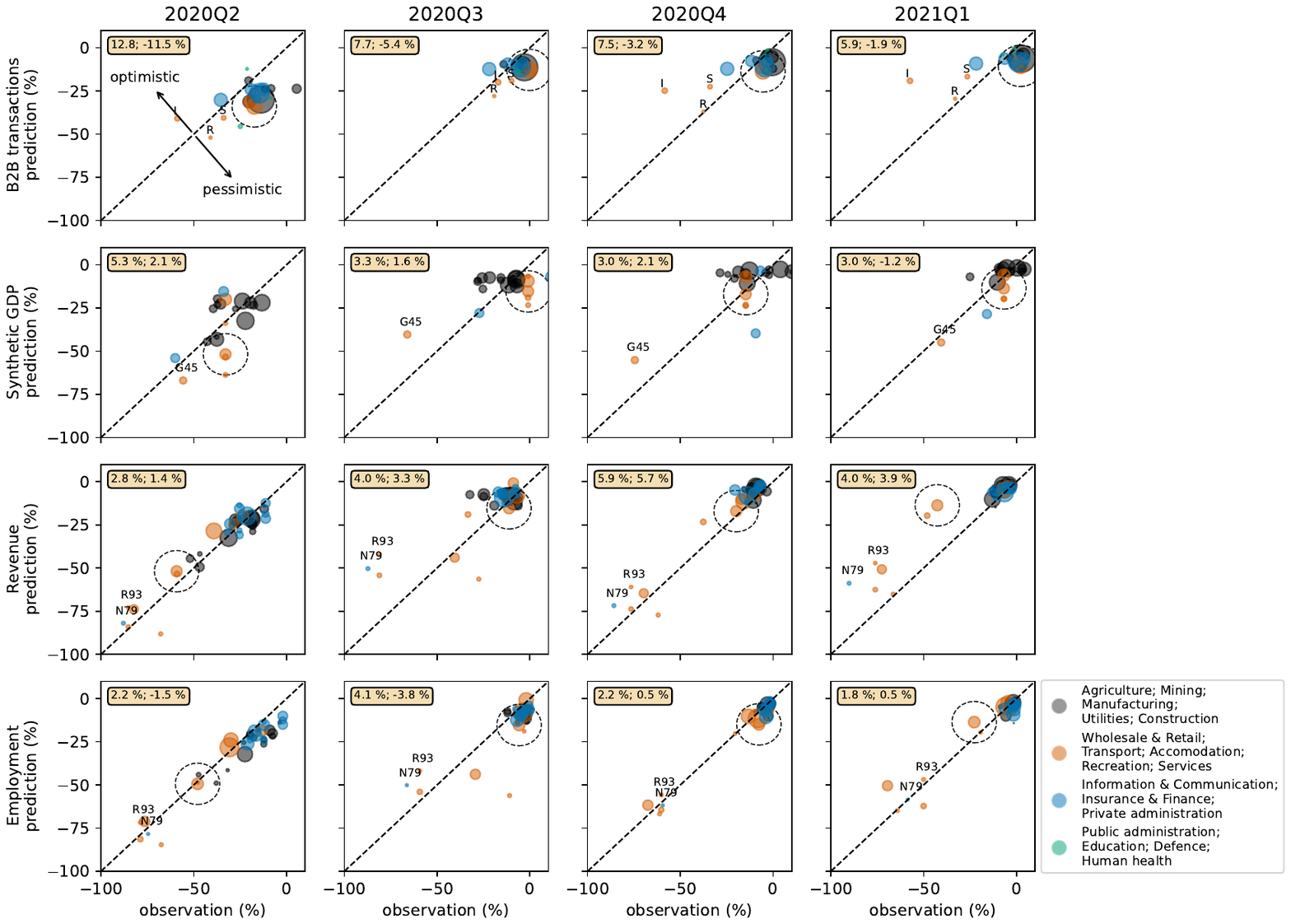}
    \caption[Comparison between the production network model projections and available data (sectoral breakdown).]{Comparison between the model projections and available data (sectoral breakdown). Using the optimal parameter values found during the sensitivity analysis (Table \ref{tab:sensitivity_global_minimum}). The dashed black circle represent transport by land, sea and air (H49, H50, H51). The marker size of each economic activity is proportional to its share in the economic indicator. In each figure, the $\text{MAE}_{\text{vw}}$ is given as a measure of accuracy while the $\text{ME}_{\text{vw}}$ is given as a measure of bias.} 
    \label{fig:plot-fit-sectors}
\end{figure}
\end{landscape}

%%%%%%%%%%%%%%%%%%%%%%%%%%%%%%%%%%%%%%%%%%%%%%%%%%%%%%%%%%%%%%%%%%%%%%%%%%%%%
%%%%%%%%%%%%%%%%%%%%%%%%%%%%%%%%%%%%%%%%%%%%%%%%%%%%%%%%%%%%%%%%%%%%%%%%%%%%%

%%%%%%%%%%%%%%%%%%%%%%%%%%%%%
%% Supplementary materials %%
%%%%%%%%%%%%%%%%%%%%%%%%%%%%%

\pagebreak
\bibliography{bibliography}

\pagebreak
\begin{appendices}

\section*{Supplementary materials}

% tables and figures must start with 'A'
\setcounter{table}{0}
\setcounter{figure}{0}
\renewcommand{\thetable}{A\arabic{table}}
\renewcommand{\thefigure}{A\arabic{figure}}

\begin{table}[!th]
    \caption[Aggregation of NACE Rev. 2 in 21 economic activities.]{Aggregation of NACE Rev. 2 in 21 economic activities.}
    \centering
    \footnotesize
    {\renewcommand{\arraystretch}{1.10}
    \begin{tabular}{>{\raggedright\arraybackslash}p{1.1cm}>{\raggedright\arraybackslash}p{7.0cm}}
        \toprule
        \textbf{Code} & \textbf{Name} \\
        \midrule
        A & Agriculture, forestry and fishing \\ 
        B & Mining and quarrying \\ 
        C & Manufacturing \\ 
        D & Electricity, gas, steam, and air conditioning supply \\ 
        E & Water supply, sewerage, waste management and remediation \\ 
        F & Construction \\ 
        G & Wholesale and retail trade \\ 
        H & Transport and storage \\ 
        I & Accommodation and food service \\ 
        J & Information and communication \\ 
        K & Finance and insurance \\ 
        L & Real estate \\ 
        M & Professional, scientific and technical activities \\ 
        N & Administration and support services \\ 
        O & Public administration and defense \\ 
        P & Education \\ 
        Q & Human health and social work \\ 
        R & Arts, entertainment and recreation \\ 
        S & Other service activities \\ 
        T & Activities of households as employers \\ 
        \bottomrule
    \end{tabular}
    }
    \label{tab:NACE21}
\end{table}

\begin{table}[!ht]
    \caption[Aggregation of NACE Rev. 2 in 64 economic activities.]{Aggregation of NACE Rev. 2 in 64 economic activities.}
    \centering
    \footnotesize
    {\renewcommand{\arraystretch}{0.80}
    \begin{tabular}{>{\raggedright\arraybackslash}p{1.2cm}>{\raggedright\arraybackslash}p{10.0cm}}
        \toprule
        \textbf{Code} & \textbf{Name} \\
        \midrule
        A01 & Agriculture \\ 
        A02 & Forestry and logging \\ 
        A03 & Fishing and aquaculture \\ 
        B05-09 & Mining and quarrying \\ 
        C10-12 & Manufacture of food, beverages and tobacco products \\ 
        C13-15 & Manufacture of textiles, wearing apparel and leather \\ 
        C16 & Manufacture of wood and of products of wood and cork, except furniture \\ 
        C17 & Manufacture of paper and paper products \\ 
        C18 & Printing and reproduction of recorded media \\ 
        C19 & Manufacture of coke and refined petroleum products \\ 
        C20 & Manufacture of chemicals and chemical products \\ 
        C21 & Manufacture of basic pharmaceutical products and pharmaceutical preparations \\ 
        C22 & Manufacture of rubber and plastic products \\ 
        C23 & Manufacture of other non-metallic mineral products \\ 
        C24 & Manufacture of basic metals \\ 
        C25 & Manufacture of fabricated metal products, except machinery and equipment \\ 
        C26 & Manufacture of computer, electronic and optical products \\ 
        C27 & Manufacture of electrical equipment \\ 
        C28 & Manufacture of machinery and equipment \\ 
        C29 & Manufacture of motor vehicles, trailers and semi-trailers \\ 
        C30 & Manufacture of other transport equipment \\ 
        C31-32 & Manufacture of furniture and other manufacturing \\ 
        C33 & Repair and installation of machinery and equipment \\ 
        D35 & Electricity, gas, steam and air conditioning supply \\ 
        E36 & Water collection, treatment and supply \\ 
        E37-39 & Sewerage; Waste collection, treatment and disposal activities; material recovery \\ 
        F41-43 & Construction of buildings; Civil engineering; Specialised construction activities \\ 
        G45 & Wholesale and retail trade and repair of motor vehicles and motorcycles \\ 
        G46 & Wholesale trade, except of motor vehicles and motorcycles \\ 
        G47 & Retail trade, except of motor vehicles and motorcycles \\ 
        H49 & Land transport and transport via pipelines \\ 
        H50 & Water transport \\ 
        H51 & Air transport \\ 
        H52 & Warehousing and support activities \\ 
        H53 & Postal and courier activities \\ 
        I55-56 & Accommodation and food services \\ 
        J58 & Publishing activities \\ 
        J59-60 & Motion picture, video and television programme production, sound recording and music publishing; Programming and broadcasting activities \\ 
        J61 & Telecommunications \\ 
        J62-63 & Computer programming, consultancy, information services \\ 
        K64 & Financial services, except insurances and pension funding \\ 
        K65 & Insurance, reinsurance, and pension funding, except compulsory social security \\ 
        K66 & Activities auxiliary to financial services and insurance activities \\ 
        L68 & Real estate \\ 
        M69-70 & Legal and accounting \\ 
        M71 & Activities of head offices; management consultancy \\ 
        M72 & Scientific research and development \\ 
        M73 & Advertising and market research \\ 
        M74-75 & Other professional, scientific and technical activities; veterinary activities \\ 
        N77 & Rental and leasing activities \\ 
        N78 & Employment activities \\ 
        N79 & Travel agencies, tour operators, and other reservation services \\ 
        N80-82 & Security and investigation activities; Services to buildings and landscape activities; Office administrative, office support and other business support activities \\ 
        O84 & Public administration and defense; compulsory social security \\ 
        P85 & Education \\ 
        Q86 & Human health activities \\ 
        Q87-88 & Residential care activities; Social work activities without accommodation \\ 
        R90-92 & Creative, arts and entertainment; Libraries, archives, museums and other cultural activities; Gambling and betting \\ 
        R93 & Sports activities and amusement and recreation activities \\ 
        S94 & Activities of membership organizations \\ 
        S95 & Repair of computers and personal and household goods \\ 
        S96 & Other personal service activities \\ 
        T97-98 & Activities of households as employers of domestic personnel; Undifferentiated goods- and services-producing activities of private households for own use \\ 
        \bottomrule
    \end{tabular}
    }
    \label{tab:NACE64}
\end{table}

\clearpage
\pagebreak

\begin{table}[!ht]
    \centering
    \footnotesize
    \caption[Overview of the production network model's initial states.]{Overview of initial states ($10^6~\text{EUR}/\text{y}$), retrieved from the (Belgian) Federal Planning Bureau \cite{FPB2018}. At equilibrium, gross output and total demand are equal, $d_i(0) = x_i(0)$. The B2B demand by sector $i$ of good $j$ is equal to the intermediate consumption listed in the input-output matrix and hence $O_{ij}(0) = Z_{ij}$. The initial stock of material $i$ held in the inventory of sector $j$, $S_{ij}(0)$ is computed as $S_{ij}(0) = n_j Z_{ij}(0)$, where $n_j$ is the inventory of material $i$ by sector $j$, expressed in the number of days production can continue if stocks are not replenished (retrieved from Pichler et al. \cite{pichler2022}).}
    {\renewcommand{\arraystretch}{0.8}
    \begin{tabular}{>{\raggedright\arraybackslash}p{1.5cm}>{\centering\arraybackslash}p{1cm}p{1cm}p{1cm}p{1cm}>{\centering\arraybackslash}p{1cm}}
        \toprule
        \textbf{Activity} & $x_{i,0}$ &  $c_{i,0}$ & $f_{i,0}$ & $l_{i,0}$ & $n_j$\\
        \midrule
        A01 & 16782 & 2489 & 3363 & 491 & 32.2 \\ 
        A02 & 648 & 93 & 163 & 23 & 39.2 \\ 
        A03 & 429 & 206 & 77 & 28 & 73.4 \\ 
        B05-09 & 24251 & 25 & 9447 & 266  & 16.8\\ 
        C10-12 & 56386 & 14792 & 23096 & 4324 & 38.5  \\ 
        C13-15 & 12802 & 3979 & 6544 & 880 & 50.6 \\ 
        C16 & 4890 & 228 & 1672 & 487 & 32.2 \\ 
        C17 & 7857 & 377 & 3138 & 642 & 28.8 \\ 
        C18 & 3193 & 63 & 640 & 674 & 16.8 \\ 
        C19 & 32573 & 3435 & 15024 & 233 & 21.5 \\ 
        C20 & 62834 & 1310 & 39364 & 3669 & 39.9 \\ 
        C21 & 21378 & 1304 & 14566 & 1548 & 47.6 \\ 
        C22 & 14087 & 559 & 7425 & 1410 & 32.8 \\ 
        C23 & 8864 & 351 & 3015 & 1491 & 36.5\\ 
        C24 & 29681 & 49 & 17145 & 1975 & 49.6 \\ 
        C25 & 14444 & 246 & 7317 & 2274 & 38.5 \\ 
        C26 & 15089 & 803 & 11006 & 672 & 52 \\ 
        C27 & 10145 & 1134 & 6066 & 1007 & 46.3 \\ 
        C28 & 23306 & 223 & 18380 & 1812 & 44.2 \\ 
        C29 & 42488 & 3705 & 31168 & 1602 & 24.5 \\ 
        C30 & 4474 & 474 & 3014 & 425 & 64.4\\ 
        C31-32 & 17193 & 2909 & 12780 & 721 & 39.2 \\ 
        C33 & 8468 & 214 & 1920 & 2325 & 37.5 \\ 
        D35 & 19084 & 5719 & 4627 & 2008 & 13.1 \\ 
        E36 & 1233 & 762 & 0 & 433 & 5.7 \\ 
        E37-39 & 14748 & 1255 & 3681 & 1581 & 11.7 \\ 
        F41-43 & 68328 & 609 & 37917 & 9383 & 64.4\\ 
        G45 & 11646 & 4234 & 4285 & 3127 & 43.6\\ 
        G46 & 56373 & 6329 & 25408 & 14907 & 18.4 \\ 
        G47 & 23611 & 22494 & 1117 & 8209 & 31.8 \\ 
        H49 & 27054 & 2217 & 8686 & 5460 & 1.7  \\ 
        H50 & 5171 & 10 & 3027 & 208 & 2 \\ 
        H51 & 7891 & 572 & 2638 & 477 & 1.7 \\ 
        H52 & 31465 & 263 & 13833 & 5370 & 25.8 \\ 
        H53 & 4405 & 191 & 715 & 1487 & 1.3 \\ 
        I55-56 & 19527 & 11300 & 1693 & 4036 & 7.4 \\ 
        J58 & 6022 & 1151 & 1714 & 802 & 7 \\ 
        J59-60 & 5167 & 868 & 1500 & 755 & 11.4 \\ 
        J61 & 14003 & 4335 & 3237 & 1797 & 6 \\ 
        J62-63 & 21334 & 0 & 9662 & 5509  & 6.4 \\ 
        K64 & 20798 & 3302 & 2537 & 3898 & 9.4 \\ 
        K65 & 9448 & 4150 & 944 & 2021 & 9.7 \\ 
        K66 & 20464 & 2691 & 6250 & 3569 & 9.4  \\ 
        L68 & 46378 & 33438 & 214 & 1166 & 34.2 \\ 
        M69-70 & 20233 & 546 & 19687 & 6691 & 21.8 \\ 
        M71 & 13253 & 94 & 5477 & 2433 & 14.7\\ 
        M72 & 20054 & 0 & 18169 & 4925 & 8.4  \\ 
        M73 & 9887 & 3 & 3821 & 798 & 3.4\\ 
        M74-75 & 2779 & 397 & 322 & 241 & 8.4 \\ 
        N77 & 17691 & 2292 & 4093 & 1214 & 3.4 \\ 
        N78 & 7661 & 0 & 50 & 6943 & 3.4 \\ 
        N79 & 3225 & 2770 & 17 & 365 & 3.4 \\ 
        N80-82 & 13999 & 1714 & 2375 & 5543 & 3.4 \\ 
        O84 & 33807 & 2729 & 30292 & 23682 & 9.4 \\ 
        P85 & 27168 & 1212 & 24225 & 21167 & 4 \\ 
        Q86 & 32665 & 6366 & 22548 & 10263 & 3 \\ 
        Q87-88 & 15209 & 6653 & 8557 & 11628 & 3 \\ 
        R90-92 & 4914 & 1983 & 1886 & 1279 & 2.3\\ 
        R93 & 2869 & 961 & 786 & 622 & 2.3 \\ 
        S94 & 6231 & 115 & 3090 & 2437 & 2.3 \\ 
        S95 & 1057 & 582 & 28 & 106 & 2.3 \\ 
        S96 & 3640 & 3241 & 7 & 620 & 2.3 \\ 
        T97-98 & 425 & 425 & 0 & 425 & 9.4 \\ 
        \bottomrule
    \end{tabular}
    }
\label{tab:overview_initial_states}    
\end{table}

\begin{figure}[ht!]
    \centering
    \includegraphics[width=0.99\linewidth]{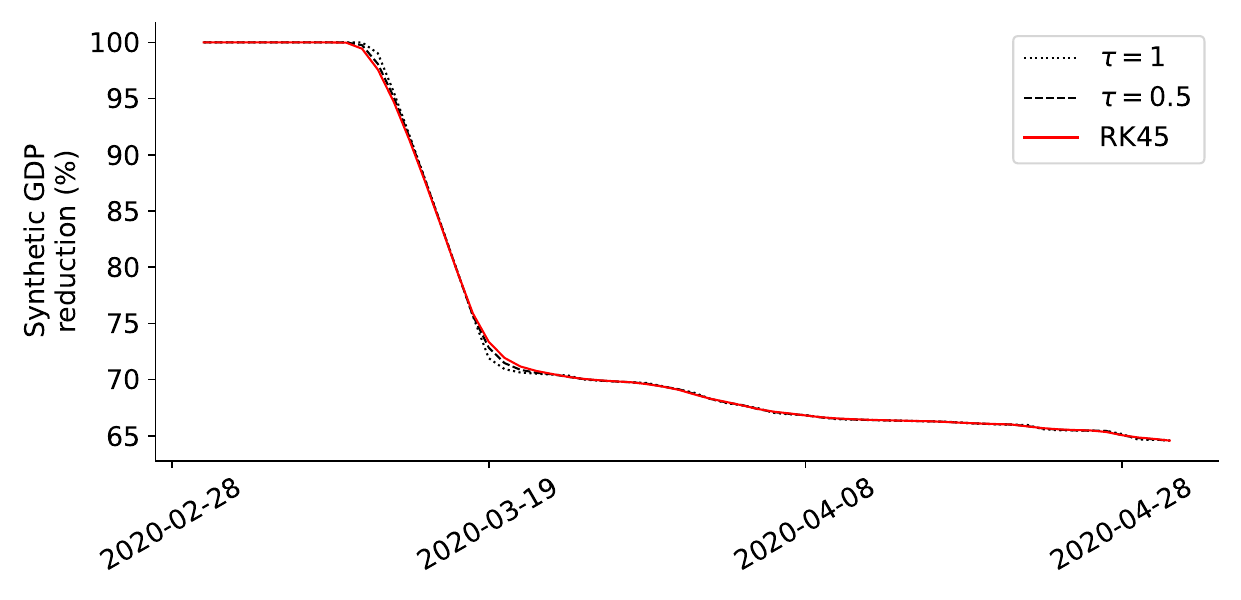}
    \caption[Impact of simulating the production network model discretely or continuously using the Runga-Kutta 45 algorithm.]{Difference in predicted aggregated gross output reduction $\big( \sum_i x_{i}(t) \big)$ when the model is simulated discretely with a step size of 0.5 and 1 days (black) versus continuously using the Runga-Kutta 45 algorithm (red). Continuous integration has a smoothing effect on sharp edges. This effect is small and switching between algorithms did not alter any of the conclusions drawn in this work.} 
    \label{fig:impact_discrete_continuous}
\end{figure}

\begin{figure}[ht!]
    \centering
    \includegraphics[width=0.99\linewidth]{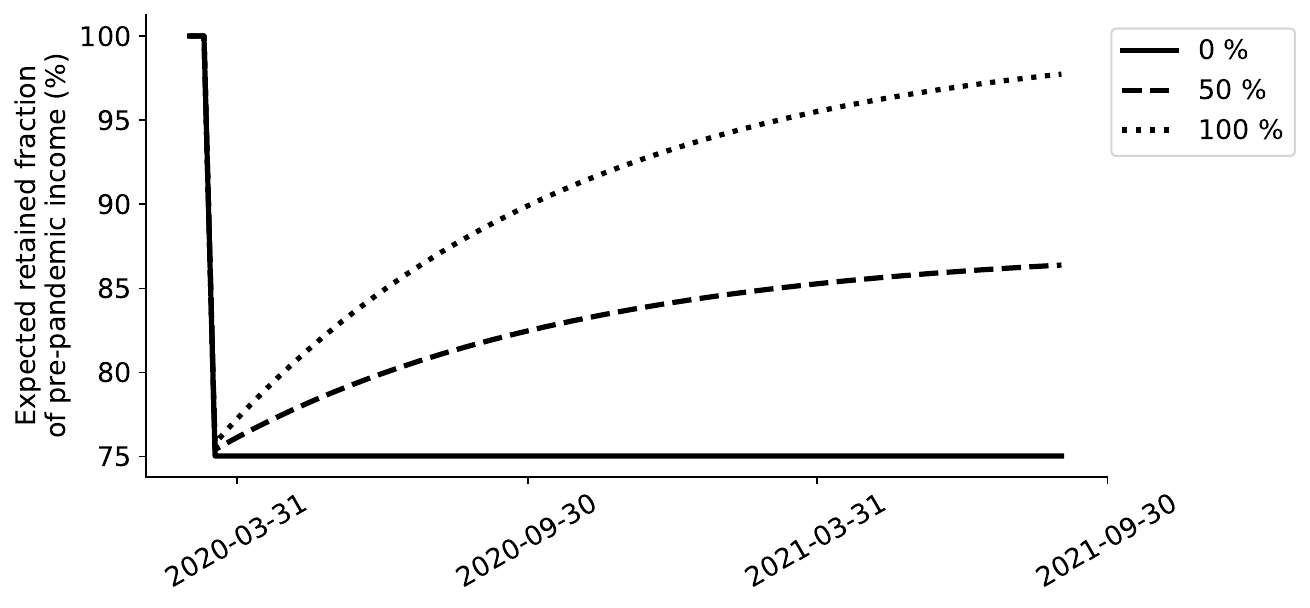}
    \caption[Expected retained long-term fraction of labor income as function of the fraction of households believing in an L-shaped recovery.]{Expected retained long-term fraction of labor income $\big(\tilde{l}_t^p\big)$ if $L=0~\%$, $L=50~\%$ or $L=100~\%$ of households believe in an L-shaped recovery.}
    \label{fig:permanent_income}
\end{figure}

\begin{figure}[ht!]
    \centering
    \includegraphics[width=0.90\linewidth]{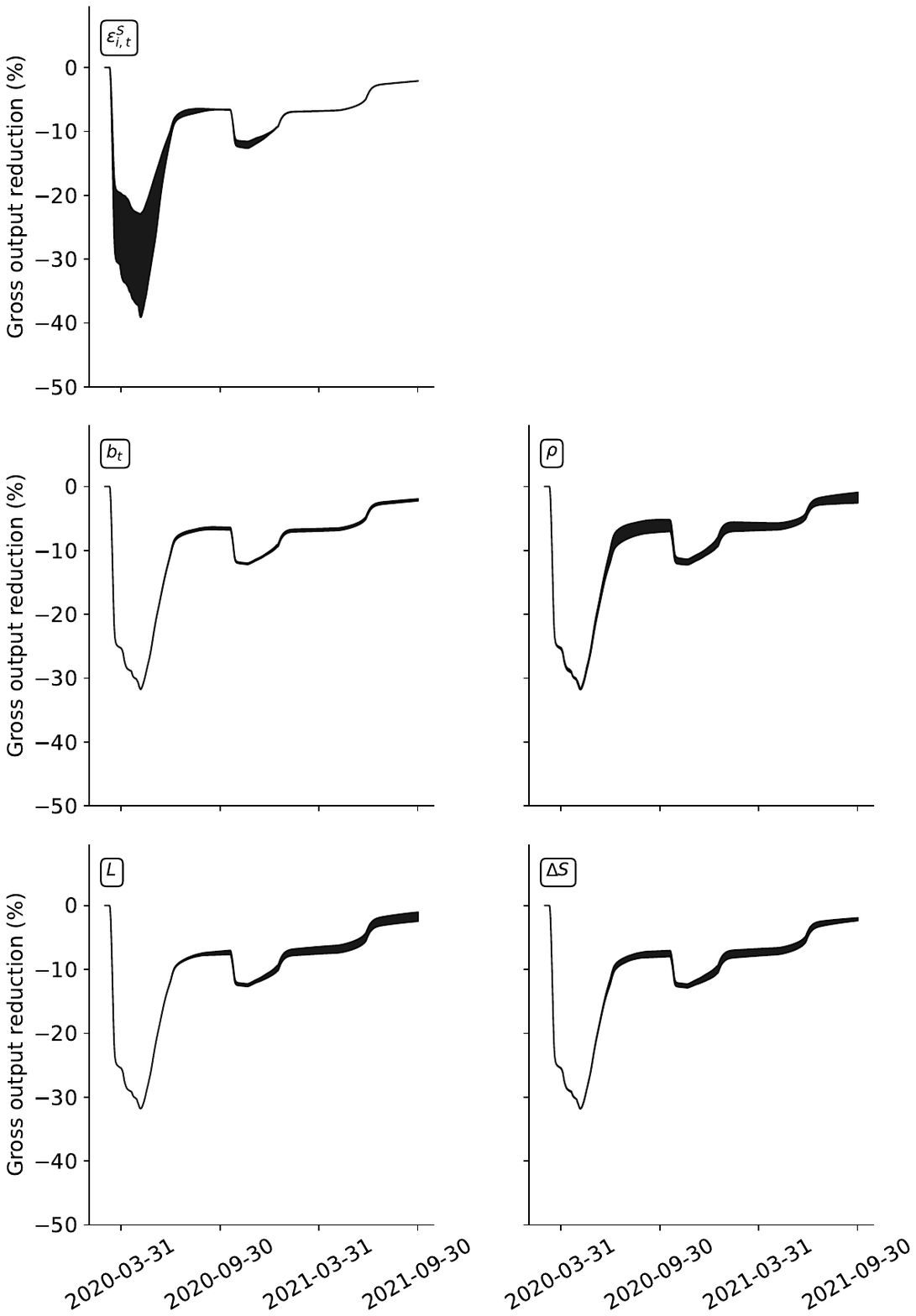}
    \caption[Sensitivity of the predicted aggregated decline in gross output to changes in the model's input parameters.]{Sensitivity of the predicted aggregated decline in gross output $\big( \sum_i x_{i}(t) \big)$ to changes in the model's input parameters. 95~\% credibility interval of 200 simulations. From top to bottom: $\kappa_{i}^S$, labor supply shocks under lockdown, sampled from $\kappa_{i}^S = \mathcal{U}(0.75,1.25)\kappa_{i}^S$. $b$, fraction of lost labor income reimbursed by the government, sampled from $b = \mathcal{U}(0.5,1)$. $\rho$, aggregate household consumption adjustment speed, sampled from $\rho = \mathcal{U}(0.1,1.0)~\text{quarters}$. $L$, fraction of households believing in an L-shaped economic recovery, sampled from $L = \mathcal{U}(0.5, 1)$. $\Delta S$, changes in the household savings rate, sampled from $\Delta S = \mathcal{U}(0.5, 1)$. Sensitivity to changes in labor supply shocks $\kappa_{i}^S$ is high while sensitivity to the other parameters is generally low.} 
    \label{fig:sensitivity_unimportant_parameters}
\end{figure}

\clearpage
\begin{figure}[th!]
    \centering
    \includegraphics[width=0.99\linewidth]{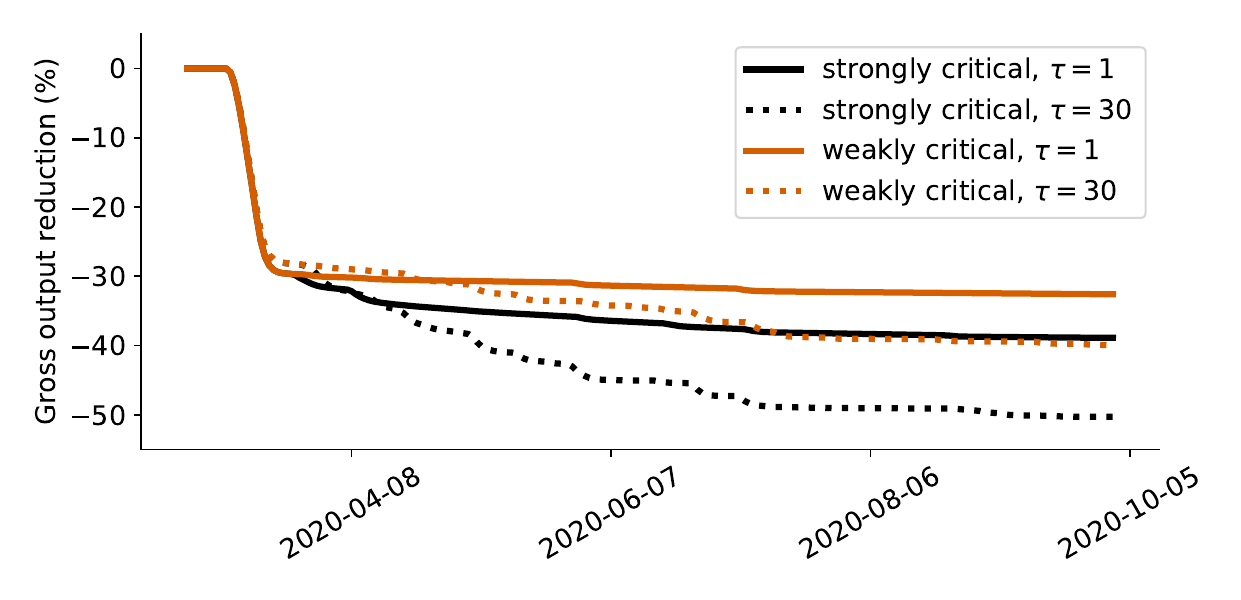}
    \caption[Predicted aggregated decline in the gross output under a prolonged first lockdown, for two production functions and for two inventory restocking speeds.]{Predicted aggregated decline in the gross output $\big( \sum_i x_{i}(t) \big)$ under a prolonged first lockdown, for two production functions and for two inventory restocking speeds. Slower inventory restocking leads to larger declines in gross output due to supply chain bottlenecks. A weakly critical PBL production function with slow inventory restocking ($\tau=30~\text{days}$) and a strongly critical PBL production function with fast inventory restocking ($\tau=1~\text{days}$) lead to similar gross output reductions under prolonged lockdown.} 
    \label{fig:sensitivity-tau}
\end{figure}

\begin{figure}[h!]
    \centering
    \includegraphics[width=\linewidth]{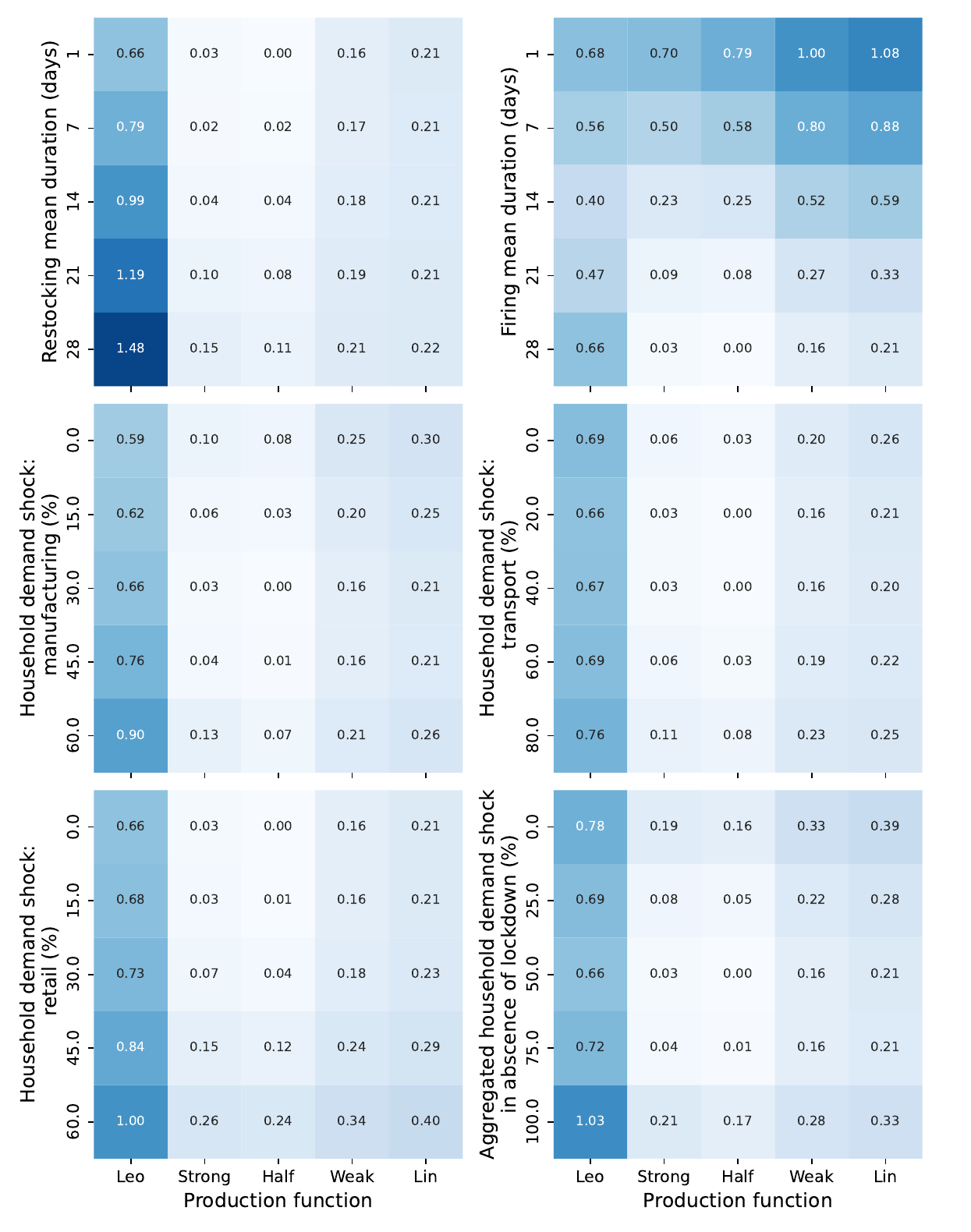}
    \caption[Two-dimensional slices of the sensitivity of the value-weighted mean average error.]{Two-dimensional slices of the sensitivity of the $\text{MAE}_{\text{vw}}$ (compared to the optimum $\Delta \text{MAE}_{\text{vw}}$).} 
    \label{fig:sensitivity_prodfunc_tau}
\end{figure}

\begin{landscape}
\begin{figure}[h!]
    \centering
    \includegraphics[width=0.85\linewidth]{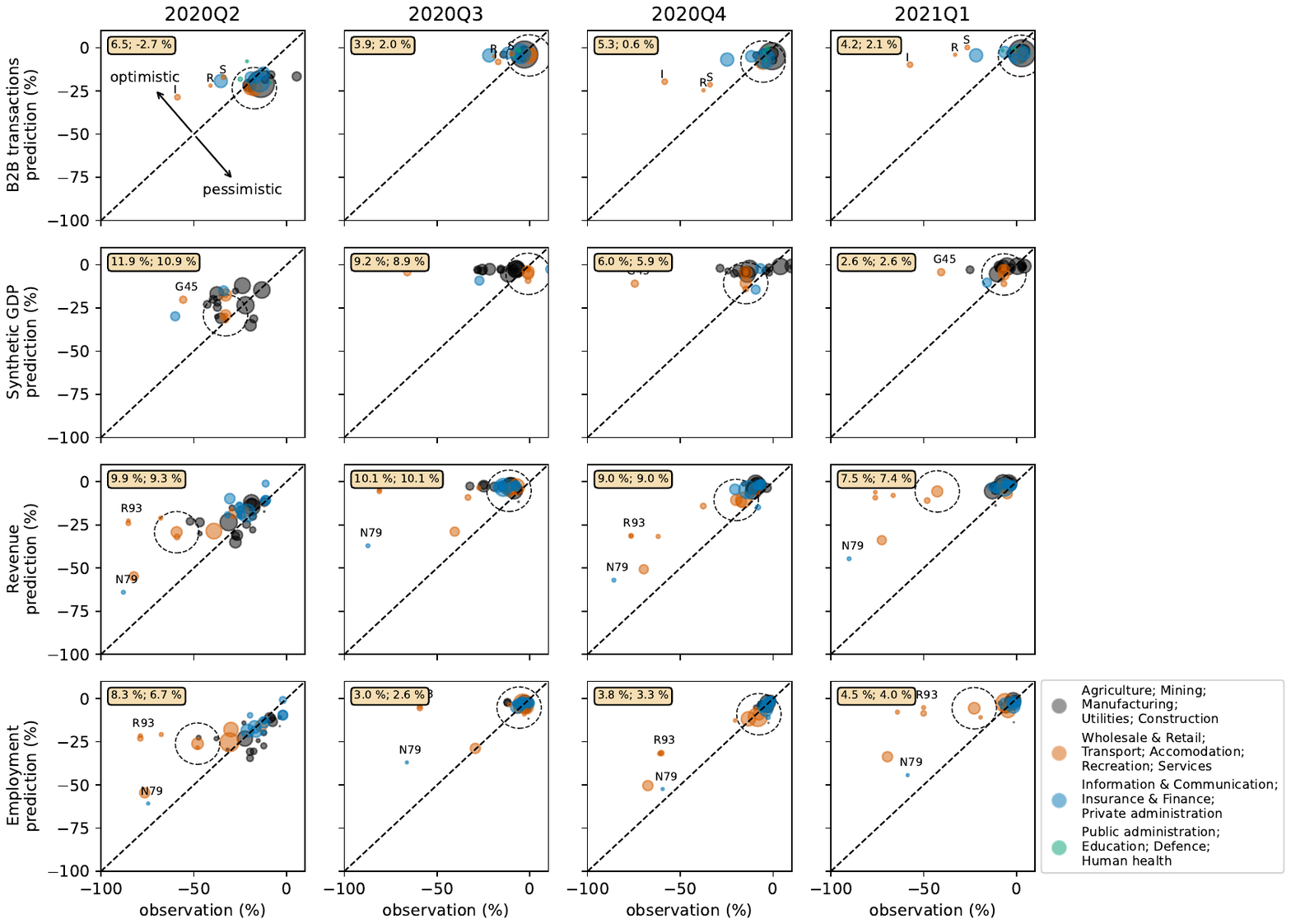}
    \caption[Comparison between the model projections and available data (sectoral breakdown). Using the parameters and shocks used by Pichler et al. (2022).]{Comparison between the model projections and available data (sectoral breakdown). Using the parameters and shocks used by Pichler et al. \citep{pichler2022}. The dashed black circle represent transport by land, sea and air (H49, H50, H51). The marker size of each economic activity is proportional to its share in the economic indicator. Results closer to the first bisector represent higher model accuracy. In each figure, the $\text{MAE}_{\text{vw}}$ is given as a measure of accuracy while the $\text{ME}_{\text{vw}}$ is given as a measure of bias.} 
    \label{fig:EPNM_comparison}
\end{figure}
\end{landscape}

\clearpage
\begin{table}[!h]
    \footnotesize
    \centering
    \caption[Overview of economic shocks caused by the COVID-19 pandemic used in the production network model.]{Overview of economic shocks caused by the COVID-19 pandemic (in \%) used in this work. labor supply shocks were obtained from the employment surveys of the ERMG \citep{ermg2021}. Sectors whose consumption happens on-site are assigned a numerical value of one while sectors without on-site consumption are assigned a value of zero.}
    {\renewcommand{\arraystretch}{0.8}
    \begin{tabular}{>{\raggedright}p{1.5cm}>{\raggedright\arraybackslash}p{1.5cm}p{1.5cm}p{1.5cm}p{1.5cm}}
        \toprule
        \textbf{Activity} & \multicolumn{2}{l}{\textbf{Labor supply}} & \multicolumn{2}{l}{\textbf{Demand}} \\ 
        ~                & \textbf{Lockdown 1} & \textbf{Lockdown 2} & \textbf{Households} & \textbf{Other} \\ 
        \midrule
        A01 & 6.5 & -5.0 & 0.0 & -27.5 \\
        A02 & 6.5 & -5.0 & 0.0 & -17.2 \\
        A03 & 6.5 & -5.0 & 0.0 & -26.7 \\
        B05-09 & 6.5 & -5.0 & 0.0 & -26.1 \\
        C10-12 & 8.5 & -3.0 & -30.0 & -26.1 \\
        C13-15 & 61.0 & -8.0 & -30.0 & -25.4 \\
        C16 & -30.0 & -5.0 & -30.0 & -35.3 \\
        C17 & -30.0 & -5.0 & -30.0 & -26.3 \\
        C18 & -28.1 & -4.0 & -30.0 & -28.8 \\
        C19 & -14.0 & -1.0 & -30.0 & -26.0 \\
        C20 & -14.0 & -1.0 & -30.0 & -25.1 \\
        C21 & -14.0 & -1.0 & -30.0 & -25.8 \\
        C22 & -19.0 & -2.0 & -30.0 & -26.0 \\
        C23 & -19.0 & -2.0 & -30.0 & -26.8 \\
        C24 & -15.0 & -6.0 & -30.0 & -25.9 \\
        C25 & -15.0 & -6.0 & -30.0 & -22.4 \\
        C26 & -13.5 & -3.0 & -30.0 & -21.7 \\
        C27 & -25.5 & -8.0 & -30.0 & -23.4 \\
        C28 & -25.5 & -8.0 & -30.0 & -22.6 \\
        C29 & -57.0 & 0.0 & -30.0 & -23.4 \\
        C30 & -57.0 & 0.0 & -30.0 & -22.4 \\
        C31-32 & -67.5 & -6.0 & -30.0 & -24.2 \\
        C33 & -28.1 & -4.0 & -30.0 & -20.8 \\
        D35 & 0.0 & 0.0 & 0.0 & -24.1 \\
        E36 & 0.0 & 0.0 & 0.0 & -25.0 \\
        E37-39 & 0.0 & 0.0 & 0.0 & -18.4 \\
        F41-43 & -43.5 & -4.0 & -30.0 & -16.8 \\
        G45 & -42.7 & -18.7 & -90.0 & -77.0 \\
        G46 & -42.7 & -18.7 & 0.0 & -5.1 \\
        G47 & -42.7 & -18.7 & 0.0 & 0.0 \\
        H49 & -61.5 & -6.0 & -50.0 & -50.0 \\
        H50 & -61.5 & -6.0 & -50.0 & -50.0 \\
        H51 & -45.0 & -1.0 & -50.0 & -50.0 \\
        H52 & -14.0 & -2.0 & 0.0 & 0.0 \\
        H53 & -61.5 & -6.0 & 0.0 & 0.0 \\
        I55-56 & -92.5 & -70.0 & -99.0 & -99.0 \\
        J58 & -16.0 & -3.0 & 0.0 & -7.6 \\
        J59-60 & -16.0 & -3.0 & 0.0 & -4.8 \\
        J61 & -16.0 & -3.0 & 0.0 & 0.0 \\
        J62-63 & -16.0 & -3.0 & 0.0 & -8.2 \\
        K64 & -2.5 & -1.0 & 0.0 & 0.0 \\
        K65 & -2.5 & -1.0 & 0.0 & 0.0 \\
        K66 & -2.5 & -1.0 & 0.0 & 0.0 \\
        L68 & -39.5 & -7.0 & 0.0 & -0.6 \\
        M69-70 & -17.0 & -4.5 & 0.0 & -0.9 \\
        M71 & -17.0 & -4.5 & 0.0 & -7.2 \\
        M72 & -17.0 & -4.5 & 0.0 & -11.4 \\
        M73 & -17.0 & -4.5 & 0.0 & 0.0 \\
        M74-75 & -17.0 & -4.5 & 0.0 & 0.0 \\
        N77 & -35.7 & -4.3 & -99.0 & -99.0 \\
        N78 & -15.5 & -5.0 & 0.0 & 0.0 \\
        N79 & -68.5 & -45.0 & -99.0 & -99.0 \\
        N80-82 & -24.0 & -6.5 & 0.0 & -2.9 \\
        O84 & 0.0 & 0.0 & 0.0 & 0.0 \\
        P85 & 0.0 & 0.0 & 0.0 & 0.0 \\
        Q86 & -40.0 & 0.0 & 0.0 & 0.0 \\
        Q87-88 & -40.0 & 0.0 & 0.0 & 0.0 \\
        R90-92 & -74.0 & -57.0 & -99.0 & -90.4 \\
        R93 & -74.0 & -57.0 & -99.0 & -99.0 \\
        S94 & -74.0 & -57.0 & -99.0 & -99.0 \\
        S95 & -28.1 & -4.0 & 0.0 & 0.0 \\
        S96 & -74.0 & -57.0 & -99.0 & -99.0 \\
        T97-98 & -97.0 & -85.0 & 0.0 & 0.0 \\
        \bottomrule
    \end{tabular}
    }
    \label{tab:overview_shocks}
\end{table}

\clearpage
\begin{table}[!ht]
    \centering
    \caption[Sectoral breakdown of the available economic time series.]{Sectoral breakdown (Table \ref{tab:NACE64}) of the available economic time series. Economic activity ``BE" refers to aggregated data for Belgium. The B2B Payment data are available for the NACE 21 classification.}
    {\footnotesize\renewcommand{\arraystretch}{0.8}
    \begin{tabular}{>{\raggedright}p{1.5cm}>{\centering}p{1.7cm}>{\centering}p{1.7cm}>{\centering}p{1.7cm}>{\centering}p{1.7cm}>{\centering\arraybackslash}p{1.0cm}}
        \toprule
        \textbf{Activity} & \textbf{Synthetic GDP} & \textbf{Employment survey} & \textbf{Revenue survey}  &  \textbf{B2B transactions} & \textbf{Total}\\ 
        \midrule
        BE & 1 & 1 & 1 & 1 & 4 \\
        A01 & ~ & 1 & 1 & 1 & 3  \\ 
        A02 & ~ & ~ & ~ & 1 & 1  \\ 
        A03 & ~ & 1 & 1 & 1 & 3\\ 
        B05-09 & ~ & ~ & ~ & 1 & 1  \\ 
        C10-12 & ~ & 1 & 1 & 1 & 3  \\ 
        C13-15 & 1 & 1 & 1 & 1 & 4  \\ 
        C16 & ~ & ~ & ~ & 1 & 1 \\ 
        C17 & 1 & 1 & 1 & 1 & 4 \\ 
        C18 & ~ & ~ & ~ & 1 & 1 \\ 
        C19 & ~ & ~ & ~ & 1 & 1 \\ 
        C20 & 1 & 1 & 1 & 1 & 4 \\ 
        C21 & ~ & 1 & 1 & 1 & 3 \\ 
        C22 & 1 & 1 & 1 & 1 & 4 \\ 
        C23 & ~ & 1 & 1 & 1 & 3 \\ 
        C24 & 1 & 1 & 1 & 1 & 4 \\ 
        C25 & 1 & 1 & 1 & 1 & 4 \\ 
        C26 & 1 & 1 & 1 & 1 & 4 \\ 
        C27 & 1 & ~ & ~ & 1 & 2 \\ 
        C28 & 1 & 1 & 1 & 1 & 4 \\ 
        C29 & 1 & ~ & ~ & 1 & 2 \\ 
        C30 & 1 & 1 & 1 & 1 & 4 \\ 
        C31-32 & ~ & 1 & 1 & 1 & 3 \\ 
        C33 & ~ & ~ & ~ & 1 & 1 \\ 
        D35 & ~ & ~ & ~ & 1 & 1 \\ 
        E36 & ~ & ~ & ~ & 1 & 1 \\ 
        E37-39 & ~ & ~ & ~ & 1 & 1\\ 
        F41-43 & 1 & 1 & 1 & 1 & 4 \\ 
        G45 & 1 & ~ & ~ & 1 & 2\\ 
        G46 & ~ & 1 & 1 & 1 & 3 \\ 
        G47 & ~ & 1 & 1 & 1 & 3 \\ 
        H49 & 1 & 1 & 1 & 1 & 4 \\ 
        H50 & 1 & ~ & ~ & 1 & 2 \\ 
        H51 & 1 & 1 & 1 & 1 & 4 \\ 
        H52 & 1 & 1 & 1 & 1 & 4\\ 
        H53 & 1 & ~ & ~ & 1 & 2 \\ 
        I55-56 & ~ & 1 & 1 & 1 & 3 \\ 
        J58 & ~ & 1 & 1 & 1 & 3 \\ 
        J59-60 & ~ & 1 & 1 & 1 & 3\\ 
        J61 & ~ & 1 & 1 & 1 & 3 \\ 
        J62-63 & 1 & 1 & 1 & 1 & 4 \\ 
        K64 & ~ & 1 & 1 & 1 & 3 \\ 
        K65 & ~ & 1 & 1 & 1 & 3\\ 
        K66 & ~ & 1 & 1 & 1 & 3 \\ 
        L68 & ~ & 1 & 1 & 1 & 3\\ 
        M69-70 & ~ & 1 & 1 & 1 & 3 \\ 
        M71 & ~ & ~ & ~ & 1 & 1 \\ 
        M72 & ~ & ~ & ~ & 1 & 1 \\ 
        M73 & ~ & ~ & ~ & 1 & 1 \\ 
        M74-75 & ~ & ~ & ~ & 1 & 1 \\ 
        N77 & 1 & ~ & ~ & 1 & 2 \\ 
        N78 & ~ & 1 & 1 & 1 & 3 \\ 
        N79 & ~ & 1 & 1 & 1 & 3 \\ 
        N80-82 & ~ & 1 & 1 & 1 & 3 \\ 
        O84 & ~ & ~ & ~ & 1 & 1 \\ 
        P85 & ~ & ~ & ~ & 1 & 1\\ 
        Q86 & ~ & ~ & ~ & 1 & 1 \\ 
        Q87-88 & ~ & ~ & ~ & 1 & 1 \\ 
        R90-92 & ~ & 1 & 1 & 1 & 3 \\ 
        R93 & ~ & 1 & 1 & 1 & 3 \\ 
        S94 & ~ & ~ & ~ & 1 & 1 \\ 
        S95 & ~ & ~ & ~ & 1 & 1 \\ 
        S96 & ~ & 1 & 1 & 1 & 3 \\ 
        T97-98 & ~ & ~ & ~ & ~ & 0 \\ 
        \midrule
        No. avail. activities & 21 & 37 & 37 & 20 & 115\\ 
        \bottomrule
    \end{tabular}
    }
    \label{tab:EPNM_overview_data}
\end{table}

\end{appendices}

\end{document}